\documentclass[11pt,british]{article}
\pdfoutput=1

\usepackage{setspace}
\usepackage{jheppub}                   
\usepackage{mathrsfs}
\usepackage[utf8]{inputenc}
\usepackage{babel}
\usepackage{booktabs}
\usepackage{microtype} 								
\SetTracking{encoding={*}, shape=sc}{0}			    
\usepackage{datetime}								
\date{\today} 										

\usepackage{amsmath}
\usepackage{amssymb}
\usepackage{amsthm}
\usepackage{mathtools}
\usepackage{esint}
\usepackage{tikz-cd}    							
\usepackage{stmaryrd}								
\SetSymbolFont{stmry}{bold}{U}{stmry}{m}{n}		    
\usepackage{slashed}								

\allowdisplaybreaks					
\numberwithin{equation}{section}	

\makeatletter
\gdef\@fpheader{\ }
\makeatother

\makeatletter 
\g@addto@macro\bfseries{\boldmath}
\makeatother

\DeclareMathOperator{\vol}{vol}					
\DeclareMathOperator{\tr}{tr}					
\DeclareMathOperator{\re}{Re}					
\DeclareMathOperator{\im}{Im}					
\DeclareMathOperator{\ad}{ad}					
\newcommand{\AdS}[1]{\text{AdS}_{#1}}			

\newcommand{\qquotient}{/\!\!/}				
\newcommand{\qqquotient}{/\!\!/\!\!/}	
\newcommand{\dd}{\mathrm{d}} 								
\newcommand{\ee}{\mathrm{e}} 								
\newcommand{\ii}{\mathrm{i}} 								
\newcommand{\Dorf}{L} 

\newcommand{\proj}[1]{\times_{#1}}							
\newcommand{\oadj}{\proj{\text{ad}}}						
\newcommand{\rep}[1]{\boldsymbol{#1}} 						
\newcommand{\repp}[2]{(\rep{#1}, \rep{#2})} 				
\newcommand{\bbZ}{\mathbb{Z}} 								
\newcommand{\bbR}{\mathbb{R}} 								
\newcommand{\bbC}{\mathbb{C}} 								
\newcommand{\bbH}{\mathbb{H}} 								
\newcommand\qqq{\qquad\quad} 								

\newcommand{\GL}[1]{\mathrm{GL}(#1)}
\newcommand{\SL}[1]{\mathrm{SL}(#1)}
\newcommand{\SU}[1]{\mathrm{SU}(#1)}
\newcommand{\Uni}[1]{\mathrm{U}(#1)}
\newcommand{\SUstar}[1]{\mathrm{SU}^{*}(#1)}

\newcommand{\Spinstar}[1]{\mathrm{Spin}^{*}(#1)}
\newcommand{\Cliff}[1]{\mathrm{Cliff}(#1)}

\newcommand{\USp}[1]{\mathrm{USp}(#1)}

\newcommand{\Ex}[1]{\mathrm{E}_{#1}}
\newcommand{\Fx}[1]{\mathrm{F}_{#1}}

\newcommand{\GDiff}{\mathrm{GDiff}}

\newcommand{\ExR}[1]{\Ex{#1}\times\mathbb{R}^+}

\newcommand{\su}[1]{\mathfrak{su}_{#1}}

\newcommand{\ex}[1]{\mathfrak{e}_{#1}}

\newcommand{\sln}[1]{\mathfrak{sl}_{#1}}
\newcommand{\gdiff}{\mathfrak{gdiff}}

\theoremstyle{definition}
\newtheorem*{defn}{Definition}
\newcommand{\MV}{\mathcal{A}_\text{V}}					
\newcommand{\MH}{\mathcal{A}_\text{H}}					
\newcommand{\MHV}{\mathcal{A}}								
\newcommand{\ModH}{\mathcal{M}_\text{H}}	

\newcommand{\HM}{H}
\newcommand{\VM}{V}

\newcommand{\HV}{HV}										
\newcommand{\ESE}{ESE}
\newcommand{\GF}{G_{\text{frame}}}
\newcommand{\norm}[1]{\left|#1\right|}

\newcommand{\trsp}{\mathrm{T}}
 
\title{The exceptional generalised geometry of supersymmetric AdS flux backgrounds}

\author[a]{Anthony Ashmore,}
\author[b]{Michela Petrini}
\author[a]{and Daniel Waldram}

\affiliation[a]{Department of Physics,
	Imperial College London,\\
	Prince Consort Road, London, SW7 2AZ, UK} 
\affiliation[b]{Sorbonne Universit\'e, UPMC Paris 06, UMR 7589,\\ LPTHE, 75005 Paris, France}

\emailAdd{a.ashmore12@imperial.ac.uk}
\emailAdd{petrini@lpthe.jussieu.fr}
\emailAdd{d.waldram@imperial.ac.uk}

\subheader{\textrm{Imperial/TP/16/DW/01}}

\abstract{We analyse generic AdS flux backgrounds preserving eight supercharges in $D=4$ and $D=5$ dimensions using exceptional generalised geometry. We show that they are described by a pair of globally defined, generalised structures, identical to those that appear for flat flux backgrounds but with different integrability conditions. We give a number of explicit examples of such ``exceptional Sasaki--Einstein'' backgrounds in type IIB supergravity and M-theory. In particular, we give the complete analysis of the generic AdS$_5$ M-theory backgrounds. We also briefly discuss the structure of the moduli space of solutions. In all cases, one structure defines a ``generalised Reeb vector'' that generates a Killing symmetry of the background corresponding to the R-symmetry of the dual field theory, and in addition encodes the generic contact structures that appear in the $D=4$ M-theory and $D=5$ type IIB cases. Finally, we investigate the relation between generalised structures and quantities in the dual field theory, showing that the central charge and R-charge of BPS wrapped-brane states are both encoded by the generalised Reeb vector, as well as discussing how volume minimisation (the dual of $a$- and $\mathcal{F}$-maximisation) is encoded. 
}

\makeatother

\begin{document}

\maketitle

\section{Introduction}

Supersymmetric AdS backgrounds are of central importance to gauge/gravity duality. In the simplest examples, corresponding to branes at conical singularities where only a top-form field strength is non-zero, they describe familiar geometries~\cite{AFHS99}, such as Sasaki--Einstein or weak-$G_2$ spaces. However, backgrounds with generic fluxes are much more complicated and at first glance have no simple geometrical description. Significant progress has been made analysing them using $G$-structures~\cite{GMPW04,GMW03,GGHPR03,GP03}, for example as means of classifying $\AdS4$ and $\AdS5$ solutions with eight supercharges in both type II theories~\cite{GMSW06} and M-theory~\cite{GMSW04c,GMPS14}. More generally one can use generalised geometry~\cite{Hitchin02,Gualtieri04,GMPT04b} to characterise the type II  backgrounds, as for example in~\cite{LT09b,MPZ06,AFPT15}. In both cases the geometry is defined by set of invariant tensors, typically only locally defined, satisfying some first-order differential equations that capture the lack of integrability of the structure in terms of the form-field flux. It is natural then to ask if there is a single notion of geometry that captures the known examples in terms of a global, integrable structure, perhaps also in a way adapted to the degrees of freedom of the dual theory. 

The answer is to use $\ExR{d(d)}$ generalised geometry~\cite{Hull07,PW08,CSW11,CSW14}, where $d=11-D$. For backgrounds with minimal supersymmetry, there is now a classification in terms of generalised special holonomy: warped supersymmetric Minkowski backgrounds are in one-to-one correspondence with spaces of particular generalised holonomy~\cite{CSW14b}. For AdS backgrounds, this was recently extended to show they are in correspondence with weak generalised special holonomy spaces~\cite{CS15}. The geometry can be characterised by a set of invariant generalised tensors, the analogues, for example, of the  $\SU{3}$-invariant two- and three-forms $\omega$ and $\Omega$ on a Calabi--Yau manifold. Using structures first considered in~\cite{GLSW09,GO12}, we showed in a recent paper~\cite{AW15} that a generic $D=4,5,6$ (warped) Minkowski background preserving eight supercharges, in type II supergravity or M-theory, defines a pair of integrable generalised structures in $\ExR{d(d)}$ generalised geometry. For $D=4,5$, one structure is naturally associated to hypermultiplets and one to vector multiplets in the Minkowski space. In particular the space of hypermultiplet structures admits a natural hyper-K\"ahler metric, while the space of vector-multiplet structures admits a very special real (if $D=5$) or special K\"ahler (if $D=4$) metric. As for a conventional $G$-structure, the generalised structures are defined by generalised tensors that are invariant under some subgroup of $\ExR{d(d)}$, and, in order to be integrable (and hence supersymmetric) must satisfy some first-order differential conditions. We should note that the formalism of ``exceptional field theory''~\cite{BP11a,BGP11,BGPW12,AGMR13,HS13} gives identical equations on the internal space to those of exceptional generalised geometry but posits not an extended tangent space but the existence of additional coordinates in spacetime. The constructions here are thus equally applicable to any such situation where a suitable enlarged spacetime can be defined.

In this paper we will give the extension of this formalism for ``exceptional Sasaki--Einstein'' geometries, that is, generic type II and M-theory AdS backgrounds in $D=4,5$ preserving eight supercharges. The generalised structures are identical to those that appear for Minkowski backgrounds, however the integrability conditions are modified in a way that depends on the cosmological constant, and is equivalent to the presence of singlet intrinsic torsion for the corresponding generalised connection~\cite{CS15}. In each case the vector-multiplet structure is defined by an invariant generalised vector which is Killing: it generates a combination of diffeomorphisms and gauge transformations that leave the background invariant, corresponding in this case to the R-symmetry of the dual field theory. By analogy to the Sasaki--Einstein case we refer to it as the ``generalised Reeb vector''. The formalism also allows one to analyse the structure of the moduli space of backgrounds. In particular we find that the space of integrable hypermultiplet structures appears as a K\"ahler slice of a hyper-K\"ahler quotient of the original space of structures, in a way closely related to the ``HK/QK correspondence'' of Haydys~\cite{Haydys08}. This mirrors the analysis of gauged $D=4,5$ supergravity~\cite{ALMTW14,LM16} precisely because the structures can be thought of as describing a rewriting of the ten- or eleven-dimensional supergravity as a $D=4,5$ theory coupled to an infinite number of hyper- and vector-multiplets~\cite{GLSW09}. 

We analyse three explicit cases to show how known supersymmetric AdS flux backgrounds appear in our formalism. For $D=5$ in type IIB, we consider the Sasaki--Einstein solutions, and also give the form of the generalised Reeb vector for the generic backgrounds in terms of spinor bilinears defined in~\cite{GMSW06}. For $D=5$ in M-theory, we give a completely general analysis, showing how the structures are defined in terms of the bilinears of~\cite{GMSW04c}, and also that the integrability conditions are satisfied. Finally, for $D=4$ in M-theory we again consider the Sasaki--Einstein solutions, and give the form of the generalised Reeb vector for the generic backgrounds in terms of bilinears of~\cite{GMPS14}. 

One striking point that emerges is the role played by the generalised Reeb vector. It is already known that, remarkably, the generic $D=5$ type IIB and $D=4$ M-theory backgrounds admit contact structures~\cite{GGPSW09,GGPSW10,GMPS14}, which encode both the central charge $a$ (or free energy $\mathcal{F}$) of the theory and the R-charges of operators dual to wrapped branes. This structure appears very naturally in the exceptional Sasaki--Einstein description: it is simply the generalised Reeb vector. As we discuss, this also leads to a very natural conjecture, following the work of~\cite{GS12}, for the generic notion of ``volume minimisation''~\cite{MSY06,MSY08}, the gravity dual of $a$- and $\mathcal{F}$-maximisation in the field theory~\cite{IW03,Jafferis10,JKPS11}.

The paper is organised as follows. We begin in section~\ref{sec:gen_struct} by reviewing the generalised structures that appear for $D=4,5$ Minkowski backgrounds preserving eight supercharges, and then recall the integrability conditions on the structures. We then move onto the main result of this paper, namely the extension of the integrability conditions for AdS backgrounds. We leave the interpretation of the conditions and a discussion of the moduli spaces of integrable structures to section~\ref{sec:sugra-mod}. We provide some concrete examples of supersymmetric AdS backgrounds in sections~\ref{sec:AdS5} and~\ref{sec:AdS4} and show they do indeed define integrable structures. In section~\ref{sec:K-physics}, we comment on the relation between vector-multiplet structures and several field theory quantities, in particular the central charge and free energy, the dimension of operators dual to wrapped branes and the dual of $a$- and $\mathcal{F}$-maximisation.  Finally, in section~\ref{sec:discuss} we finish with a short summary and discussion of areas for further work.

The notation and conventions used in this paper can be found in appendices~A and~E of~\cite{AW15}.

\section{Generalised structures for AdS}\label{sec:gen_struct}

We begin by reviewing the generalised structures that define $D=4,5$ backgrounds preserving eight supercharges. These were defined in~\cite{AW15} for Minkowski vacua, but are equally applicable to AdS. The only difference is in the integrability conditions, and one of the main results of this paper is to give the conditions relevant to AdS. We provide some concrete examples, including the case of completely general fluxes in M-theory giving an $\AdS5$ vacuum. We leave the interpretation of the conditions and a discussion of the moduli spaces of integrable structures to section~\ref{sec:sugra-mod}.

\subsection{Hyper- and vector-multiplet structures in \texorpdfstring{$\Ex{d(d)}$}{Ed(d)} generalised geometry}

We consider type II and M-theory solutions of the form $\AdS{D} \times M$, where $M$ is $(10-D)$-dimensional for type II and $(11-D)$-dimensional for M-theory. We assume the metric is a warped product
\begin{equation}
\label{eq:metric_ansatz}
  \dd s^2 = \ee^{2\Delta} \dd s^2(\AdS{D}) + \dd s^2 (M),
\end{equation}
where $\Delta$ is a scalar function on $M$. We take $m$ to be the inverse AdS radius, so that the Ricci tensor is normalised to $R_{\mu\nu}=-(D-1)m^2 g_{\mu\nu}$, where $g$ is the metric on $\AdS{D}$, and the cosmological constant is $\Lambda=-\tfrac{1}{2}(D-1)(D-2)m^2$. As in~\cite{AW15}, we allow generic fluxes compatible with the AdS symmetry of the external spacetime and use the string frame metric for type II solutions.

As shown in~\cite{AW15}, a generic background preserving eight supercharges is completely characterised by a pair of generalised $G$-structures in exceptional generalised geometry. These structures were first defined in~\cite{GLSW09}, in the context of type II theories. The generalised tangent bundle $E$ in exceptional generalised geometry~\cite{Hull07,PW08} admits an action of $\ExR{d(d)}$. We can then define a generalised frame bundle $\tilde{F}$ for $E$ as an $\ExR{d(d)}$ principal bundle. There is also a generalised Lie derivative~\cite{PW08,BGGP12,CSW11} which encodes the infinitesimal symmetries, diffeomorphisms and gauge transformations, of the supergravity theory, and one can use it to define generalised torsion and the analogue of the Levi-Civita connection~\cite{CSW11,CSW14}. Generalised tensors are defined as sections of vector bundles transforming in some representation of $\ExR{d(d)}$. A generalised $G$-structure is then defined by a set of generalised tensors that are invariant under the action of a subgroup $G\subset\Ex{d(d)}$. Equivalently, it is a choice of $G$ principal sub-bundle of the generalised frame bundle $\tilde{P}_G\subset\tilde{F}$. The notion of an integrable generalised structure as one with vanishing intrinsic torsion then follows in analogy to the conventional case~\cite{CSW14b}. 

The pairs of structures that appear for $\mathcal{N}=2$, $D=4$ and $\mathcal{N}=1$, $D=5$ backgrounds were named hypermultiplet and vector-multiplet structures, or \HM{} and \VM{} structures for short, since they are associated to hyper- and vector-multiplet scalar degrees of freedom in $D$ dimensions. The relevant structure groups defined by the \HM{} and \VM{} structures are summarised in table~\ref{tab:HVstructures}. 
\begin{table}
\centering
\begin{tabular}{@{}rllll@{}} 
\toprule
& $\GF$ & \HM{} structure & \VM{} structure & \HV{} structure \\
\midrule
$D=4$ & $\ExR{7(7)}$ & $\Spinstar{12}$ & $\Ex{6(2)}$ & $\SU6$ \\
$D=5$ & $\ExR{6(6)}$ & $\SUstar6$ & $\Fx{4(4)}$ & $\USp6$ \\
\bottomrule 
\end{tabular}
\protect\caption{The generalised $G$-structures with $G\subset\Ex{7(7)}$ and $G\subset\Ex{6(6)}$ that define eight-supercharge backgrounds in $D=4$ and $D=5$ respectively.}
\label{tab:HVstructures}
\end{table}
The hypermultiplet structure is defined by a triplet of sections of a weighted adjoint bundle 
\begin{align}
\label{eq:J-def}
   \text{\HM{} structure :} & \qquad 
   J_\alpha \in \Gamma(\ad\tilde{F}\otimes(\det T^*M)^{1/2})
   \qquad \alpha=1,2,3, 
\end{align}
which define a highest weight $\su2$ subalgebra of $\ex{d(d)}$ and are normalised using the $\ex{d(d)}$ Killing form such that \begin{equation}\label{eq:J_conditions}
   [J_\alpha, J_\beta]=2\kappa\epsilon_{\alpha\beta\gamma}J_\gamma,\qqq
   \tr(J_\alpha J_\beta)=-\kappa^2 \delta_{\alpha\beta}.
\end{equation}
Similarly, the vector-multiplet or \VM{} structure is defined by a section of the generalised tangent bundle $E$
\begin{align}
\label{eq:T-def}
   \text{\VM{} structure :} & \qquad K \in \Gamma(E) , 
\end{align}
which has a positive norm with respect to the $\Ex{7(7)}$ quartic invariant $q(K)>0$ or the $\Ex{6(6)}$ cubic invariant $c(K)>0$.\footnote{Recall that for $\Ex{7(7)}$ there is a symmetric quartic invariant $q(V_1,V_2,V_3,V_4)$ and a symplectic invariant $s(V_1,V_2)$ and for $\Ex{6(6)}$ a symmetric cubic invariant $c(V_1,V_2,V_3)$. We use the shorthand $q(V)=q(V,V,V,V)$ and $c(V)=c(V,V,V)$.} In $D=4$, one can use the quartic invariant as a Hitchin function to define a second invariant tensor $\hat{K}$ and combine the two into a complex object 
\begin{equation}
\label{eq:X-def}
   X = K + \ii \hat{K} . 
\end{equation}
Explicitly, $\hat{K}$ is defined by the relation 
\begin{equation}
\label{eq:s-q-relation}
   s(V,\hat{K})=2 q(K)^{-1/2}q(V,K,K,K).
\end{equation}
for arbitrary $V\in\Gamma(E)$. 

Finally the pair of structures $\{J_\alpha,K\}$ define an \HV{} structure if they are compatible, that is, if they satisfy the conditions
\begin{equation}\label{eq:compatibility}
 \text{\HV{} structure :} \qqq
 J_\alpha\cdot K=0, \qqq 
 \tr(J_\alpha J_\beta) = \begin{dcases}
-2\sqrt{q(K)}\delta_{\alpha\beta}& D=4, \\
-c(K)\delta_{\alpha\beta} & D=5,
\end{dcases}
\end{equation}
where for $D=4$ from~\eqref{eq:s-q-relation} we also have $\sqrt{q(K)}=\frac{1}{2}s(K,\hat{K})$. 

Given a pair of globally defined spinors on $M$, one can construct  ``untwisted'' structures $\{\tilde{J}_\alpha,\tilde{K}\}$ in terms of spinor bilinears. The full structures include the potentials for the appropriate form fields and are given by the exponentiated adjoint action on the untwisted objects~\cite{AW15}, thus in M-theory we have
\begin{equation}
\label{eq:M-twist}
   J_\alpha = \ee^{A+\tilde{A}} \tilde{J}_\alpha \ee^{-A-\tilde{A}} , \qquad
   K = \ee^{A+\tilde{A}} \tilde{K} , 
\end{equation}
where $A$ is the three-form potential and $\tilde{A}$ is the dual six-form potential, and for type IIB 
\begin{equation}
\label{eq:IIB-twist}
   J_\alpha = \ee^{B^i+C} \tilde{J}_\alpha \ee^{-B^i-C} , \qquad
   K = \ee^{B^i+C} \tilde{K} , 
\end{equation}
where $B^i$ is the $\SL{2;\bbR}$ doublet of two-form potentials and $C$ is the four-form potential. In this case one also needs to include dressing by the IIB axion and dilaton, as described in appendix~E of~\cite{AW15}. Since these transformations are in $\Ex{d(d)}$, the algebra, normalisation and compatibility conditions~\eqref{eq:J_conditions} and~\eqref{eq:compatibility} can be checked using either the twisted or untwisted objects.

\subsection{Exceptional Sasaki--Einstein geometry}
\label{sec:ESE}

We now describe the integrability conditions on the \HV{} structure for the case of a supersymmetric AdS background preserving eight supercharges. As discussed in~\cite{CSW14b,CS15}, the difference from the Minkowski case is that there is a constant singlet component of the generalised intrinsic torsion, resulting in a background with weak generalised holonomy. This leads to a simple modification of the Minkowski conditions given in~\cite{AW15}.

Recall that the space of \HM{} structures has a natural hyper-K\"ahler cone geometry and admits a triplet of moment maps for the action of the generalised diffeomorphism group $\GDiff$, that is, the diffeomorphism and gauge transformation symmetries of the underlying supergravity theory. Infinitesimal transformations are generated by the generalised Lie derivative $\Dorf_V$ and so are parameterised by generalised vectors $V\in\Gamma(E)$. The moment maps for a given element in $\gdiff$ parameterised by $V$ are given by   
\begin{equation}
\label{eq:mm-def}
   \mu_\alpha(V) = 
      -\tfrac{1}{2}\epsilon_{\alpha\beta\gamma}\int_M
          \tr(J_\beta \Dorf_V J_\gamma).
\end{equation}
For Minkowski backgrounds, supersymmetry implied that the moment maps vanished. For AdS backgrounds they take a fixed non-zero value. Let us define the real functions
\begin{align}
  D=4: \qqq
  \gamma(V) &= 2\int_M q(K)^{-1/2}q(V,K,K,K) ,  \label{eq:cosmo_d4}\\
  D=5: \qqq
  \gamma(V) &= \int_M c(V,K,K) \label{eq:cosmo_d5} .
\end{align}   
Note that the first definition can also be written in terms of $\hat{K}$ using~\eqref{eq:s-q-relation}. 

We can then define the exceptional generalised geometry analogue of a Sasaki--Einstein structure, corresponding to an AdS background with generic fluxes. We have 
\begin{defn}
An \emph{exceptional Sasaki--Einstein (\ESE{}) structure} is an \HV{} structure $\{J_\alpha,K\}$ satisfying 
\begin{align}
   \mu_\alpha(V) &= \lambda_\alpha \gamma(V) \qquad \forall V\in\Gamma(E) , 
   \label{eq:mm-gen} \\
   \Dorf_K K &= 0 , 
   \label{eq:VM_diff} \\
   \Dorf_K J_\alpha 
      &= \epsilon_{\alpha\beta\gamma} \lambda_\beta J_\gamma, 
      \qqq \Dorf_{\hat{K}}J_\alpha=0,
   \label{eq:HV_diff}
\end{align}
where $\gamma(V)$ is given by~\eqref{eq:cosmo_d4} and~\eqref{eq:cosmo_d5}, and $\lambda_\alpha$ are real constants related to the inverse AdS radius by $\norm{\lambda}=2m$ for $D=4$ and $\norm{\lambda}=3m$ for $D=5$, where $\norm{\lambda}^2=\lambda_1^2+\lambda_2^2+\lambda_3^2$. The second condition in~\eqref{eq:HV_diff} is relevant only for $D=4$.
\end{defn}
\noindent
The integrability condition for the vector-multiplet structure~\eqref{eq:VM_diff} is unchanged from the Minkowski case. As shown in~\cite{AW15}, for $D=4$ this is equivalent to $\Dorf_X \bar{X} = 0$, as $\Dorf_X X$ vanishes identically. The other two conditions now have right-hand sides, determined by the singlet torsion. Note that the third condition~\eqref{eq:HV_diff} simply states that the action of $\Dorf_K$ is equivalent to an $\SU2$ rotation of the $J_\alpha$. Note also that this condition is consistent with the moment map conditions when taking $V=K$ (and $V=\hat{K}$ in $D=4$). As shown in appendix~\ref{app:lemma}, for ESE spaces, the second compatibility constraint in~\eqref{eq:compatibility} is actually a consequence of the integrability conditions. 

Recall that for $D=4$ there is a global $\Uni1$ R-symmetry that acts on $X$, taking $X\to X'=\ee^{\ii\alpha}X$. Strictly, one should write the condition~\eqref{eq:HV_diff} replacing $K$ with $\re X'$ and $\hat{K}$ with $\im X'$. However, the point is that one can always choose a gauge where the condition takes the form~\eqref{eq:HV_diff}. In a similar way one can use the $\SU2$ global R-symmetry to set $\lambda_{1,2}=0$. (The only unbroken part of the R-symmetry is then a $\Uni1$ preserving $\lambda_3$.) The conditions~\eqref{eq:mm-def} can then be written as 
\begin{equation}
\label{eq:mm} 
   \mu_3(V) = \lambda_3 \gamma(V) , \qqq
   \mu_+(V) = 0 , \qqq \forall V\in\Gamma(E) ,  
\end{equation}
while the conditions~\eqref{eq:HV_diff} read
\begin{align}
D=4:\qquad
\Dorf_K J_+ &= \ii \lambda_3 J_+ ,\qquad \Dorf_{\hat{K}} J_+ = 0, \label{eq:HV_diff_4}\\
D=5:\qquad
\Dorf_K J_+ &= \ii \lambda_3 J_+ \label{eq:HV_diff_5}. 
\end{align}   
These are the forms we will use when checking the integrability for various examples. 

We can immediately deduce various properties from the integrability conditions. The first is that the \ESE{} space is generalised Einstein. Recall that the \HV{} structure $\{J_\alpha,K\}$ determines the generalised metric $G$ that encodes the supergravity degrees of freedom on $M$. Given a generalised metric one can construct a unique generalised Ricci tensor following~\cite{CSW11}. Using the generalised intrinsic torsion of the \ESE{} background, which we discuss in section~\ref{sec:int-tor}, and the results of~\cite{CS15}, we find that the generalised Ricci tensor satisfies the generalised Einstein equation\footnote{We are using $R_{MN}=R^0_{MN}+\frac{1}{d_E}G_{MN}R$, where $R^0$ and $R$ are the generalised tensors defined in~\cite{CSW11}.}
\begin{equation}
   R_{MN} = \frac{(D-1)(D-2)}{d_E} m^2 G_{MN} , 
\end{equation}
where $M$ and $N$ are indices running over the dual generalised tangent space $E^*$ and $d_E$ is the dimension of $E$.

Next we note that since $\Dorf_KK=0$ and $\Dorf_K J_\alpha$ is equal to an $\SU2$ R-symmetry rotation, which simply rotates the Killing spinors but leaves the supergravity degrees of freedom unchanged, we can conclude that $\Dorf_K G = 0$ and so
\begin{equation}
   \Dorf_K G = 0 \quad \Leftrightarrow 
      \quad \text{$K$ is a generalised Killing vector} , 
\end{equation}
as is also the case for Minkowski backgrounds. Note that for the $D=4$ solutions, $\hat{K}$ is also generalised Killing. Let us decompose $K$ into vector and form components as in~\cite{AW15}, 
\begin{equation}
   K = \begin{cases}
          \xi + \omega + \sigma + \tau & \text{M theory}, \\
          \xi + \lambda^i + \rho + \sigma^i + \tau 
             & \text{type IIB}, 
          \end{cases}
\end{equation}
where $\xi$ is the vector component. The generalised Killing vector condition in M-theory is equivalent to
\begin{equation}
\label{eq:gen-Killing-m}
\begin{aligned}
   \mathcal{L}_\xi g &=0 , & &&
   \mathcal{L}_\xi A - \dd \omega &= 0 , & &&
   \mathcal{L}_\xi \tilde{A} - \dd \sigma 
       + \tfrac{1}{2} \dd\omega\wedge A &= 0 , 
\end{aligned}
\end{equation}
where $A$ is the three-form potential and $\tilde{A}$ is the dual six-form potential. In type IIB it is equivalent to
\begin{equation}
\label{eq:gen-Killing-IIB}
\begin{aligned}
   \mathcal{L}_\xi g &=0 , & &&
   \mathcal{L}_\xi C &= \dd \rho 
      - \tfrac{1}{2} \epsilon_{ij}\dd\lambda^i\wedge B^j  ,  \\
   \mathcal{L}_\xi B^i &= \dd \lambda^i , & && 
   \mathcal{L}_\xi\tilde{B}^i &= \dd \sigma^i 
      + \tfrac{1}{2}\dd\lambda^i\wedge C - \tfrac{1}{2}\dd\rho\wedge B^i +\tfrac{1}{12}B^i \wedge \epsilon_{kl} B^k \wedge \dd\lambda^l , 
\end{aligned}
\end{equation}
where $B^i$ is the $\SL{2;\bbR}$ doublet of two-form potentials, $\tilde{B}^i$ are their six-form duals and $C$ is the four-form potential. In each case, the form components give compensating gauge transformations so that the field strengths ($F=\dd A$ etc.) are invariant under the diffeomorphism generated by $\xi$. We immediately see that if $\xi=0$ then all the form components are closed and hence $\Dorf_K$ acting on \emph{any} generalised tensor vanishes. However, this is in contradiction with the condition~\eqref{eq:HV_diff}. Hence we conclude that $\xi$ is non-zero and the solution admits a Killing vector that also preserves all the fluxes. Furthermore from~\eqref{eq:HV_diff} we see that it generates the unbroken $\Uni1\subset\SU2$ R-symmetry. On Sasaki--Einstein spaces this vector is known as the Reeb vector. Thus we are led to define
\begin{defn}
We call $K$ the \emph{generalised Reeb vector} of the exceptional Sasaki--Einstein geometry, noting that its vector component $\xi\in\Gamma(TM)$ is necessarily non-vanishing.
\end{defn}
The fact that $K$ is generalised Killing means that, in the untwisted frame where there are no potentials in the generalised metric, the corresponding ``twisted'' generalised Lie derivative must reduce to just a conventional Lie derivative, that is 
\begin{equation}
\label{eq:Lie-xi}
   \hat{\Dorf}_{\tilde{K}} = \mathcal{L}_{\xi} , 
\end{equation}
where $\xi$ is the vector component of $\tilde{K}$ (and hence also of $K$). 
Acting on an arbitrary untwisted generalised tensor $\tilde{\alpha}$, the twisted generalised Lie derivative takes the form 
\begin{equation}
\label{eq:twisted-Dorf}
   \hat{\Dorf}_{\tilde{V}} \tilde{\alpha} 
       = \mathcal{L}_v \tilde{\alpha} - \tilde{R} \cdot \tilde{\alpha} , 
\end{equation}
where $\tilde{R}$ is a tensor in the adjoint representation of $\Ex{d(d)}$, $\tilde{R}\cdot\tilde{\alpha}$ is the adjoint action, $v$ is the vector component of $\tilde{V}$, $\mathcal{L}_v$ is the conventional Lie derivative and 
\begin{equation}
\label{eq:tildeR}
   \tilde{R} = \begin{cases}
           \dd\tilde{\omega} - \imath_{\tilde{v}}F
           + \dd\tilde{\sigma} -\imath_{\tilde{v}}\tilde{F} 
           + \tilde{\omega}\wedge F & \text{for M-theory} \\
            \dd\tilde{\lambda}^i - \imath_{\tilde{v}}F^{i}
           + \dd\tilde{\rho} - \imath_{\tilde{v}}F 
           - \epsilon_{ij}\tilde{\lambda}^{i}\wedge F^{j}
           + \dd\tilde{\sigma}^i 
           + \tilde{\lambda}^{i}\wedge F
           - \tilde{\rho}\wedge F^{i} & \text{for type IIB}
        \end{cases} .
\end{equation}
The condition~\eqref{eq:Lie-xi} thus means that the corresponding tensor $\tilde{R}$ vanishes. The conditions~\eqref{eq:VM_diff} and~\eqref{eq:HV_diff} can then be written as  
\begin{equation}
\label{eq:Lie-JK}
   \mathcal{L}_{\xi} \tilde{J}_\alpha 
        = \epsilon_{\alpha\beta\gamma}\lambda_\beta \tilde{J}_\gamma , \qquad
   \mathcal{L}_{\xi} \tilde{K} = 0 . 
\end{equation}
In what follows it will sometimes be simpler when checking solutions to use these forms of the conditions. 

Finally, we note that there is a consistency condition on $K$ implied by the moment map conditions~\eqref{eq:mm-gen}. Strictly, there is a kernel in the map $\Dorf_{\bullet} : \Gamma(E)\rightarrow \gdiff$, meaning that two different generalised vectors can generate the same generalised diffeomorphism. In other words, we have $\Dorf_V=\Dorf_{V+\Delta}$, which holds if
\begin{equation}
\label{eq:Delta-conds}
   \Delta = \begin{cases}
          \omega + \sigma + \tau 
          & \text{with $\dd\omega=\dd\sigma=0$ in M-theory}, \\
          \lambda^i + \rho + \sigma^i + \tau 
          & \text{with $\dd\lambda^i=\dd\rho=\dd\sigma^i=0$ in type IIB}. 
          \end{cases}
\end{equation}
Thus for the conditions~\eqref{eq:mm-gen} to make sense we need
\begin{equation}
\label{eq:gamma-delta}
   \gamma(\Delta) = 0 , 
\end{equation}
which is a differential condition on $K$. In fact it is implied by the conditions~\eqref{eq:VM_diff} and~\eqref{eq:HV_diff}. Note first that these conditions are satisfied by both $K$ and $K+\Delta$. As we have already mentioned, substituting~\eqref{eq:HV_diff} into the expression for the moment maps~\eqref{eq:mm-def} gives
\begin{equation}
\label{eq:mm-K}
   \mu_\alpha(K) = \lambda_\alpha \int_M \kappa^2
      = \lambda_\alpha \gamma(K) ,
\end{equation}
where the second equality follows from the second of the compatibility conditions~\eqref{eq:compatibility}. From the homogeneity of $q$ and $c$, we note that upon taking the functional derivative, where $M$ runs over all the components of the generalised vector, we have
\begin{equation}
   \int_M V^M \frac{\delta \gamma(K)}{\delta K^M}
      = (D-2) \gamma(V) . 
\end{equation}
Then note that, using $\mu_\alpha(K+\Delta)=\mu_\alpha(K)$ and~\eqref{eq:mm-K}, we have
\begin{equation}
   0 = \int_M \Delta^M \frac{\delta \mu_\alpha(K)}{\delta K^M}
      = \lambda_\alpha (D-2)\gamma(\Delta) , 
\end{equation}
and hence indeed $\gamma(\Delta)=0$. Note that this derivation did not use the moment map conditions~\eqref{eq:mm-gen} themselves, only the conditions~\eqref{eq:VM_diff} and~\eqref{eq:HV_diff} involving $\Dorf_K$. 

Finally, in the $D=4$ case $\hat{K}$ is also generalised Killing. However, from the condition $\gamma(\Delta)=0$ and~\eqref{eq:s-q-relation}, we have
\begin{equation}
   \gamma(\tau) = \int_M s(\tau,\hat{K}) = 0 , 
\end{equation}
for all $\tau$ for both type IIB and M-theory. From the form of the symplectic invariant given in~\cite{AW15}, this implies that the vector component of $\hat{K}$ vanishes. Since $\hat{K}$ is Killing this means 
\begin{equation}
   \Dorf_{\hat{K}} ( \text{anything} ) = 0, 
\end{equation}
or in other words, $\hat{K}$ is in the kernel of the map $\Dorf_{\bullet} : \Gamma(E)\rightarrow \gdiff$, satisfying the same conditions as $\Delta$ in~\eqref{eq:Delta-conds}. As such, it generates a trivial generalised diffeomorphism and hence the generalised metric is not invariant under a second symmetry; only $K$ generates a non-trivial transformation. 

\subsection{Generalised intrinsic torsion}
\label{sec:int-tor}

As conjectured in~\cite{CSW14b} and proven in~\cite{CS15}, the Killing spinor equations for generic AdS flux backgrounds preserving eight supercharges are in one-to-one correspondence with \HV{} structures with constant singlet generalised intrinsic torsion.\footnote{Strictly for $D=4$ only the $\mathcal{N}=1$ case was considered in~\cite{CS15}. However, combined with the comments about $\mathcal{N}=2$ in~\cite{CSW14b}, the results of~\cite{CS15} are sufficient to prove that for $\mathcal{N}=2$ there is a constant singlet torsion transforming in a triplet of $\SU2$.} In each case the non-zero torsion was in the $\repp{3}{1}$ of $\SU2\times G$, where $G$ is the \HV{} structure group, which breaks the $\SU2$ R-symmetry to $\Uni1$. These were called spaces with weak generalised special holonomy, in analogy with conventional $G$-structures. This is in constrast to Minkowski backgrounds where all components of the intrinsic torsion vanished. Note that there are no singlets in the generalised intrinsic torsion for $D=6$, giving the standard result that there are no $\mathcal{N}=(1,0)$ AdS solutions in six dimensions. 

In order to prove that our conditions~\eqref{eq:mm-gen},~\eqref{eq:VM_diff} and~\eqref{eq:HV_diff} are equivalent to the conditions for supersymmetry, we need to check that they indeed admit a constant non-zero singlet in the $\repp{3}{1}$ component of the intrinsic torsion. To do this we can simply repeat the calculations of~\cite{AW15}. One immediately notes that the $\repp{3}{1}$ component appears in the moment maps and $\Dorf_KJ_\alpha$, but not $\Dorf_KK$. This explains why the $\Dorf_KK=0$ condition is unchanged from the Minkowski case. By definition, the right-hand side of~\eqref{eq:HV_diff} is a constant singlet in $\repp{3}{1}$ as it is a constant linear combination of $J_\alpha$. Consistency with the moment maps then implies~\eqref{eq:mm-gen} for $V=K$. This proves that the integrability conditions are indeed equivalent to the Killing spinor equations.

\section{Gauged supergravity and moduli spaces}
\label{sec:sugra-mod}

\subsection{Integrability conditions from gauged supergravity}
\label{sec:sugra-int}

As stressed in~\cite{GLSW09,AW15}, the infinite-dimensional spaces $\MH$ and $\MV$ of hyper- and vector-multiplet structures correspond to a rewriting of the ten- or eleven-dimensional supergravity theory so that only eight supercharges are manifest~\cite{DN86}. The local Lorentz symmetry is broken and the fields of the theory can be reorganised into $\mathcal{N}=2$, $D=4$ or $\mathcal{N}=1$, $D=5$ multiplets  without making a Kaluza--Klein truncation. One can then interpret the integrability conditions in terms of conventional gauged $D=4$ or $D=5$ supergravity with an infinite-dimensional gauging by $\GDiff$. The general conditions for supersymmetric vacua have been given in~\cite{HLV09,LST12,LM16}, and it was shown in~\cite{AW15} that for Minkowski backgrounds these conditions are precisely the integrability conditions on the generalised structures. 

Let us now briefly show that the same is true for the AdS backgrounds. Following~\cite{LST12}, a generic gauged $\mathcal{N}=2$, $D=4$ theory admits an AdS vacua provided
\begin{equation}
\label{eq:4dvacua-conds}
\Theta_\Lambda^\lambda \mu_{\alpha,\lambda}=-\tfrac{1}{2}
 \ee^{K^\text{v}/2} \Omega_{\Lambda\Sigma}\im(\hat{\mu}\bar{X}^\Sigma)a_\alpha,\qquad
 \bar{X}^\Lambda \hat{\Theta}^{\hat{\lambda}}_\Lambda \hat{k}^i_{\hat{\lambda}}=0 , \qquad
 X^\Lambda \Theta_\Lambda^\lambda k_\lambda^u = c a_\alpha (\xi^\alpha)^u ,
\end{equation}
where $\norm{\hat{\mu}}\propto m$, $a_\alpha$ is unit-norm vector parametrising $S^2$, $K^{\text{v}}$ is the K\"ahler potential and $\Omega_{\Lambda\Sigma}$ the symplectic structure on the space of vector multiplets $\MV$. We have written the last condition not on the quaternionic-K\"ahler space, but on the corresponding hyper-K\"ahler cone. Any Killing vector preserving the quaternionic-K\"ahler structure on the base lifts to a vector that rotates the three complex structures on the cone. Thus $(\xi^\alpha)^u$ are the three vectors generating the $\su2$ action the cone, normalised such that $\xi^\alpha\cdot \xi^\beta=\delta^{\alpha\beta}$. There is a consistency condition between the first and third conditions that arises from the identity $k_\lambda\cdot \xi^\alpha=-2\mu_{\alpha,\lambda}$~\cite{WKV00,WRV01b}. This is the same consistency we already noted for the integrability conditions~\eqref{eq:HV_diff} and~\eqref{eq:VM_diff}. Contracting the third expression in~\eqref{eq:4dvacua-conds} with $\xi^\alpha$ and the first expression with $X^\Lambda$, we find 
\begin{equation}
   c = \ee^{K^\text{v}/2} \Omega_{\Lambda\Sigma}
      X^\Lambda \im(\hat{\mu}\bar{X}^\Sigma) . 
\end{equation}
We can then choose $\hat{\mu}$ to be real using the $\Uni1$ action on $X$. Using the identifications between terms in the $\mathcal{N}=2$ expressions and the \HM{} and \VM{} geometries discussed in~\cite{AW15}, we see that, using~\eqref{eq:s-q-relation} and for real $\hat{\mu}$, the three conditions in~\eqref{eq:4dvacua-conds}  exactly match~\eqref{eq:mm-gen},~\eqref{eq:VM_diff} and~\eqref{eq:HV_diff} respectively. Explicitly we can identify
\begin{equation}
\begin{aligned}
   V^\Lambda \Theta_\Lambda^\lambda \mu_{\alpha,\lambda} &= \mu_\alpha(V) , \\
   \Omega_{\Lambda\Sigma}V^\Lambda\im(\hat{\mu}\bar{X}^\Sigma)a_\alpha
         &= a_\alpha \Omega(V,\hat{K}) = a_\alpha \gamma(V) , \\
   \bar{X}^\Lambda \hat{\Theta}^{\hat{\lambda}}_\Lambda \hat{k}_{\hat{\lambda}}
       &= \Dorf_{\bar{X}}X , 
\end{aligned}
\end{equation}
and $\ee^{-K^\text{v}}=\ii\Omega(X,\bar{X})$. While acting on the section-valued functions $J_\alpha$, we have
\begin{equation}
\begin{aligned}
   a_\beta \xi^\beta(J_\alpha) &= -\epsilon_{\alpha\beta\gamma}a_\beta J_\gamma , \\
   X^\Lambda \Theta_\Lambda^\lambda k_\lambda(J_\alpha) &= \Dorf_X J_\alpha . 
\end{aligned}
\end{equation}
It is straightforward to see that conditions in $D=5$ can similarly be matched to the gauged supergravity expressions for AdS vacua given in~\cite{LST12}. 

\subsection{Moduli spaces of ESE backgrounds}
\label{sec:moduli}

We now turn to analysing the structure of the moduli space of exceptional Sasaki--Einstein backgrounds satisfying the integrability conditions~\eqref{eq:mm-gen}--\eqref{eq:HV_diff}. Given the relation to gauged supergravity discussed above, we can use known results on the form of the moduli space of AdS vacua in these theories~\cite{ALMTW14,LM16}. For example, for $\mathcal{N}=2$, $D=4$ gauged supergravity, it was shown in~\cite{ALMTW14} that the vector-multiplet moduli space is a real subspace of the local special K\"ahler manifold~$\MV/\bbC^*$, while the hypermultiplet moduli space is a K\"ahler submanifold of the quaternionic manifold $\MH/\bbH^*$, at least in the so called ``minimal'' solution. More generally, the combined moduli space is no longer a product. 

In fact, the situation here is more complicated because we have to impose the compatibility conditions~\eqref{eq:compatibility} between the \HM{} and \VM{} structures. This means that even before imposing the integrability conditions, the space $\MHV$ of \HV{} structures is not actually a product $\MV\times\MH$. Nonetheless, as described in~\cite{AW15}, if we drop the normalisation part of the compatibility condition, we can view $\MHV$ as a fibration of a hyper-K\"ahler cone space over a special K\"ahler space (or vice versa). The same structure arises for $D=5$ but now we have a hyper-K\"ahler cone over a very special real manifold (or vice versa). 

Focussing for definiteness on $D=5$, though an analogous analysis applies to $D=4$, we can use this fibration picture to analyse the form of the moduli space. Let us first fix a generalised Reeb vector $K\in\MV$ satisfying the integrability condition $\Dorf_KK=0$. We can now consider the space of \HM{} structures $\MH^K\subset\MH$ compatible with the fixed $K$, that is 
\begin{equation}
   \MH^K = \left\{ J_\alpha\in \MH : J_\alpha\cdot K = 0 \right\}. 
\end{equation}
We can drop the normalisation condition $\kappa^2=c(K)$ since, as we show in appendix~\ref{app:lemma}, it is a consequence of the supersymmetry conditions. At each point on the manifold $M$, the space of possible $J_\alpha$ is given by the hyper-K\"ahler cone 
\begin{equation}
   W = \bbR^+ \times \Fx{4(4)} / \USp6 , \qquad 
   \text{$W/\bbH^*$ is a Wolf space} , 
\end{equation}
and in complete analogy to the construction of $\MH$ we find that the infinite-dimensional space $\MH^K$ is itself a hyper-K\"ahler cone. We are now left with imposing the remaining two supersymmetry conditions
\begin{equation}
   \mu_\alpha(V) = \lambda_\alpha \gamma(V) , \qquad 
   \Dorf_K J_\alpha = \epsilon_{\alpha\beta\gamma} \lambda_\beta J_\gamma . 
\end{equation}

We would like to have geometrical interpretations of both conditions. Recall first that since $\MH^K$ is a hyper-K\"ahler cone it admits a free $\SU2$ action generated by a triple of vectors $\xi^\alpha\in\Gamma(T\MH^K)$. The action of $\GDiff$ is triholomorphic (it preserves all three symplectic structures) and is generated by a vector $\rho_V\in\Gamma(T\MH^K)$ for each $V\in E$. By definition, acting on the $J_\alpha$ we have 
\begin{equation}
   \rho_V(J_\alpha) = \Dorf_V J_\alpha , \qquad
   \xi^\alpha(J_\beta) = \epsilon_{\alpha\beta\gamma} J_\gamma . 
\end{equation}
Because of the ``source'' term $\lambda_\alpha \gamma(V)$ in the moment maps, only a subgroup $\Uni1\subset\SU2$ of transformations  leave the moment map conditions invariant. This group is generated by $r=\lambda_\alpha \xi^\alpha$ and preserves one linear combination of complex structures $I=\lambda_\alpha I^\alpha$ on $\MH^K$. Restricting to $V=K$, the vector $\rho_K$ generates a one-dimensional subgroup $G_K\subset\GDiff$ corresponding to the generalised diffeomorphisms generated by $K$. As shown in~\cite{AW15}, these two actions commute. 

We can now interpret the condition~\eqref{eq:HV_diff} as a vector equation
\begin{equation}
\label{eq:fp-cond}
   \rho_K - r = 0 , 
\end{equation}
that is, it restricts us to points on $\MH^K$ that are fixed points of a combined action of $G_K$ and $\Uni1$. (Note that generically we expect that fixed points will only exist for certain choices of $K$ satisfying $\Dorf_KK=0$.) We define 
\begin{equation}
   \mathcal{N}_{\text{H}} = \bigl\{ p \in \MH^K : \rho_K(p)-r(p)=0 \bigr\} .
\end{equation}
Since both $\rho_K$ and $r$ preserve the complex structure $I$, both are real holomorphic vectors and hence $\mathcal{N}_{\text{H}}$ is a K\"ahler subspace of $\MH^K$ with respect to $I$. 

Let us now turn to the moment maps. We would like to view them as defining a hyper-K\"ahler quotient. Thought of as single map  $\mu:\MH^K\to\gdiff^*\times\bbR^3$, for AdS backgrounds, the level set defined by~\eqref{eq:mm-gen} is $\mu^{-1}(\Lambda_\alpha)$, where the element $\Lambda_\alpha\in\gdiff^*\times\bbR^3$ is given by the functional derivative  $\Lambda_\alpha=\lambda_\alpha\delta\gamma/\delta V$. But since $\gamma(V)$ depends on $K$ we see that it is not invariant under the full generalised diffeomorphism group. A hyper-K\"{a}hler quotient is well defined only on a level set that is invariant under the action of the quotient group. However, we can define a subgroup of generalised diffeomorphisms $\GDiff_K\subset\GDiff$ as those that leave $K$ invariant, that is the stabiliser group, 
\begin{equation}
   \GDiff_K = \{ \Phi\in\GDiff : \Phi\cdot K = K \} ,
\end{equation}
so that infinitesimally, $V$ parametrises an element of the corresponding algebra $\gdiff_K$ if $\Dorf_V K=0$. Since $\Dorf_KK=0$ note that $G_K\subset\GDiff_K$.  For a fixed $K$, any two \HM{} structures related by an element of $\GDiff_K$ are equivalent. If we restrict to the subgroup $\GDiff_K$, then we can view the moment maps as a hyper-K\"ahler quotient.\footnote{The one caveat is that the conditions~\eqref{eq:mm-gen} are satisfied for arbitrary $V$ parametrising all of $\gdiff$ not just $V$ with $\Dorf_VK=0$ parametrising $\gdiff_K$. Thus we need to be sure the conditions arising from the moment maps with restricted $V$, together with the other supersymmetry conditions~\eqref{eq:VM_diff} and~\eqref{eq:HV_diff}, are sufficient. Although we have not found a general proof, we can see this is true in a number of explicit examples. This is not surprising, since, as shown in~\cite{AW15}, the moment maps only constrain a relatively small independent component $\repp{2}{6}$ of the intrinsic torsion.} Since the moment map conditions break the $\SU2$ action to $\Uni1$, although the quotient space is by definition hyper-K\"ahler, it is not a hyper-K\"ahler cone, that is, there is no longer an underlying quaternionic-K\"ahler space. 

Combining the quotient with the fixed-point conditions~\eqref{eq:fp-cond} we then have two possibilities: either take a quotient and then impose~\eqref{eq:fp-cond} or impose~\eqref{eq:fp-cond} and then take a quotient. Doing the latter we note that the fixed-point condition already imposes that we are on a K\"ahler subspace, so there is no notion of a hyper-K\"ahler quotient. However, we show in appendix~\ref{app:lemma} that, restricting to $\GDiff_K$ on $\mathcal{N}_{\text{H}}$, two of the moment map conditions are identically satisfied. Thus we are actually only taking a \emph{symplectic} quotient with a moment map given by $\mu(V)=\lambda_\alpha\mu_\alpha(V)$. Thus we have the diagram
\begin{equation}
\label{eq:quotients}
\begin{tikzcd}[row sep=large, column sep=huge]
\MH^K \arrow{r}{\rho_K-r=0} \arrow{d}{\text{HK quotient}} 
& \mathcal{N}_{\text{H}} \arrow{d}{\text{sympl.~quotient}} \\
\ModH' \arrow{r}{r'=0} & \ModH  
\end{tikzcd}  
\end{equation}
where $\ModH'=\MH^K\qqquotient\GDiff_K$ is a hyper-K\"ahler manifold, and the final moduli space 
\begin{equation}
   \ModH = \mathcal{N}_{\text{H}} \qquotient \GDiff_K \qquad \text{is K\"ahler}.   
\end{equation}
The vector $r'$ in~\eqref{eq:quotients} generates the $\Uni1$ action on the quotient space $\ModH'$. Since the action of $\rho_K$ is modded out on the quotient space, it is trivial and so the condition becomes just $r'=0$. However, since $r'$ is still real holomorphic with respect to $I$, we see that going via $\ModH'$, the space $\ModH$ is again K\"ahler. One caveat to taking the hyper-K\"ahler quotient first is that there might be additional solutions to $r'=0$. Since $r$ is freely acting,  we have $r'=0$ whenever there is a generalised diffeomorphism such that $\Dorf_VJ_\alpha=\epsilon_{\alpha\beta\gamma}\lambda_\beta J_\gamma$. However, since $\Dorf_VK=0$ as $V\in \gdiff_K$, we see that such $V$ are generalised Killing vectors. Thus, provided $K$ is the only generalised Killing vector, we can take either path in the diagram~\eqref{eq:quotients}. 

We can slightly refine the construction to make a connection to the ``HK/QK correspondence'' of Haydys~\cite{Haydys08}, which physically is related to the c-map. This also helps the analysis in the case where there are fixed points. Given $V$ satisfying $\Dorf_VK=0$, acting on any generalised tensor $\alpha$ we have 
\begin{equation}
   [\Dorf_V,\Dorf_K]\alpha = \Dorf_{\Dorf_VK}\alpha = 0 .
\end{equation}
Thus $G_K$ is in the centre of $\GDiff_K$ and as such is a normal subgroup. Thus we can define the quotient group $\GDiff_K^0=\GDiff_K/G_K$ and write $\GDiff_K$ as a semi-direct product
\begin{equation}
   \GDiff_K = G_K \rtimes \GDiff_K^0 . 
\end{equation}
We can then perform the hyper-K\"ahler quotient in two stages: first by the action of $G_K$ and then by $\GDiff_K^0$, as described in symplectic case, for example, in~\cite{MMPR98}. We can then add one more level to the diagram~\eqref{eq:quotients}
\begin{equation}
\label{eq:quotients2}
\begin{tikzcd}[row sep=large, column sep=huge]
\MH^K \arrow{r}{\rho_K-r=0} \arrow{d}{\bullet\qqquotient G_K} 
& \mathcal{N}_{\text{H}} \arrow{d}{\bullet\qquotient G_K} \\
\mathcal{P}_{\text{H}} \arrow{r}{r'=0}
\arrow{d}{\bullet\qqquotient\GDiff_K^0}
& \mathcal{Q}_{\text{H}} 
\arrow{d}{\bullet\qquotient\GDiff_K^0} \\
\ModH' \arrow{r}{r'=0} & \ModH  
\end{tikzcd}  
\end{equation}
Consider the path through the diagram with two commuting Abelian actions on $\MH^K$ given by $G_K$ and $\Uni1\subset\SU2$, with the latter preserving only one linear combination of the three complex structures. This is exactly the set up that appears in the HK/QK correspondence~\cite{Haydys08}: the hyper-K\"ahler manifold is $\mathcal{P}_{\text{H}}$ while the quaternionic-K\"ahler manifold is $\MH^K/\bbH^*$.

\section{\texorpdfstring{$\AdS5$}{AdS5} backgrounds as ESE spaces}\label{sec:AdS5}

We now discuss the structure of exceptional Sasaki--Einstein (ESE) backgrounds for $\AdS5$. The generic flux backgrounds for type IIB were analysed in~\cite{GMSW06}, and for M-theory in~\cite{GMSW04c}. Here we first show how the standard type IIB Sasaki--Einstein reduction with five-form flux embeds as an ESE background, and comment on how this extends to the generic case. We then give the ESE form of the generic M-theory background, showing explicitly how the integrability conditions reproduce those given in~\cite{GMSW04c}. 

\subsection{Sasaki--Einstein in type IIB}

Backgrounds of the form $\text{AdS}_{5}\times M$, where the five-dimensional space $M$ is Sasaki--Einstein and there is a non-trivial self-dual five-form flux, are supersymmetric solutions of type IIB supergravity preserving at least eight supercharges~\cite{KW98}. The metric is a product of the form \eqref{eq:metric_ansatz} with $D=5$ and a constant warp factor, which we take to be zero. Five-dimensional Sasaki--Einstein spaces admit a nowhere-vanishing vector field $\xi$, known as the Reeb vector and a pair two-forms $\Omega$ and $\omega$, that together define an $\SU2\subset\GL{5;\bbR}$ structure (for a review see for example~\cite{BG08,Sparks11}). They satisfy the algebraic conditions 
\begin{equation}\label{eq:SE5_alg}
\Omega\wedge\bar{\Omega}=2\omega\wedge\omega,\qquad\imath_{\xi}\Omega=\imath_{\xi}\omega=0,\qquad\imath_{\xi}\sigma=1,
\end{equation}
where $\sigma$ is the one-form constructed from $\xi$ by lowering the index with the metric (that is $\xi=\sigma^\sharp$). In addition one has the differential conditions 
\begin{equation}
\label{eq:SE5-integ}
\dd\sigma=2m\omega , \qquad \dd\Omega=3\ii m\sigma\wedge\Omega, 
\end{equation}
where $m$ is the inverse $\AdS5$ radius, usually normalised to $m=1$. Such a compactification is supersymmetric provided there is a five-form flux given by 
\begin{equation}
   \dd C=F=4m\vol_{5},
\end{equation}
where $\vol_{5}=-\tfrac{1}{2}\sigma\wedge\omega\wedge\omega$. 

Note that these conditions imply that the Reeb vector $\xi$ is a Killing vector that preserves $\sigma$ and $\omega$, but rotates $\Omega$ by a phase
\begin{equation}
\mathcal{L}_{\xi}\sigma=\mathcal{L}_{\xi}\omega=\mathcal{L}_{\xi}g=0,\qquad\mathcal{L}_{\xi}\Omega=3\ii m \Omega.
\end{equation}
The rotation of $\Omega$ corresponds to the R-symmetry of the solution. In what follows we also need the (transverse) complex structure
\begin{equation}
I_{\phantom{m}n}^{m}=-\omega_{\phantom{m}n}^{m}
    =\tfrac{\ii}{4}(\bar{\Omega}{}^{mp}\Omega_{np}-\Omega^{mp}\bar{\Omega}_{np}),
\end{equation}
which satisfies $I_{\phantom{p}m}^{p}\Omega_{pn}=\ii\Omega_{mn}$.


The Sasaki--Einstein geometry defines an ``unwtwisted'' \HV{} structure given by an \HM{} structure invariant under $\SUstar6$
\begin{equation}
\begin{aligned}
   \tilde{J}_{+} &= \tfrac{1}{2}\kappa u^{i}\Omega 
       + \tfrac{1}{2}\kappa v^{i}\Omega^{\sharp}, \\
   \tilde{J}_{3} &= \tfrac{1}{2}\kappa I
      + \tfrac{1}{2}\kappa\hat{\tau}^{i}_{\phantom{i}j}
      + \tfrac{1}{8}\kappa\Omega^{\sharp}\wedge\bar{\Omega}^{\sharp}
      - \tfrac{1}{8}\kappa\Omega\wedge\bar{\Omega},
\end{aligned}
\end{equation}
where $u^i=(-\ii,1)^i$, $v^i=(-1,-\ii)^i$, $\hat{\tau}$ is given in terms of the second Pauli matrix $\hat{\tau}=-\ii\sigma_{2}$, and the $\Ex{6(6)}$-invariant volume is $\kappa^2=\vol_5$. The \VM{} structure invariant under $\Fx{4(4)}$ is given by 
\begin{equation}
   \tilde{K}=\xi-\sigma\wedge\omega .
\end{equation}
Using the adjoint action and the $\ex{6(6)}$ Killing form in~\cite{AW15}, one can check that $\tilde{J}_\alpha$ satisfy the $\su2$ algebra and are correctly normalised as in~\eqref{eq:J_conditions}, while using the cubic invariant from~\cite{AW15} and the algebraic conditions~\eqref{eq:SE5_alg}, one can check that $\tilde{K}$ and $\tilde{J}_\alpha$ satisfy the compatibility conditions~\eqref{eq:compatibility}, so that together $\{J_\alpha,K\}$ define a $\USp{6}$ structure. The full ``twisted'' structures include the four-form potential $C$ as in~\eqref{eq:IIB-twist}, however, in what follows, it will actually be easier to work with the untwisted structures and use the twisted generalised Lie derivative in the differential conditions.


Let us now see how the integrability conditions on $\sigma$, $\omega$, $\Omega$ and $F$ arise. We turn first to the moment map conditions~\eqref{eq:mm}. Let $\tilde{V}$ be an untwisted generalised vector. Using the untwisted $\tilde{K}$, we see that the function~\eqref{eq:cosmo_d5} takes the form 
\begin{equation}
   \gamma(\tilde{V}) 
      = \tfrac{1}{3}\int_M \imath_{\tilde{v}}\sigma\vol_{5}
        + \omega\wedge\tilde{\rho} ,
\end{equation}
where $\tilde{v}$ and $\tilde{\rho}$ are the vector and three-form components of $\tilde{V}$. As the moment map condition must hold for an arbitrary generalised vector, we can consider each component of $\tilde{V}$ in turn. We begin with the $\tilde{\rho}$ components of $\mu_3$:
\begin{equation}
\begin{split}
   \mu_{3}(\tilde{\rho})-\lambda_3\gamma(\tilde{\rho}) 
   &= - \tfrac{1}{8}\int_{M}\kappa^2 
      (\Omega^\sharp\wedge\bar{\Omega}^\sharp)\lrcorner\dd\tilde{\rho}
      - \tfrac{1}{3}\lambda_3 \int_M\tilde{\rho}\wedge\omega \\
& =\int_{M}\tfrac{1}{2}\dd\tilde{\rho}\wedge\sigma - \tfrac{1}{3}\lambda_3 \tilde{\rho}\wedge\omega,
\end{split}
\end{equation}
which vanishes for $\dd\sigma=\tfrac{2}{3}\lambda_3\omega$. Next we consider the $\mu_{+}$ condition, which gives
\begin{equation}
\begin{split}\mu_{+}(\tilde{V}) & \propto\int_{M}\kappa^{2}\Omega^{\sharp}\lrcorner\dd(\tilde{\lambda}^{1}+\ii\tilde{\lambda}^{2})\\
& \propto\int_{M}(\Omega^{\sharp}\lrcorner\vol_{5})\wedge\dd(\tilde{\lambda}^{1}+\ii\tilde{\lambda}^{2})\\
& \propto\int_{M}\dd(\sigma\wedge\Omega)\wedge(\tilde{\lambda}^{1}+\ii\tilde{\lambda}^{2}).
\end{split}
\end{equation}
Using $\dd\sigma\propto\omega$ from the previous condition, this vanishes for $\sigma\wedge\dd\Omega=0$. Finally we have the $\tilde{v}$ components of $\mu_3$:
\begin{equation}
\begin{split}
   \mu_{3}(\tilde{v})-\lambda_3\gamma(\tilde{v}) & =\tfrac{1}{8}\int_{M}
      \ii \mathcal{L}_{\tilde{v}} \Omega\wedge\sigma\wedge\bar{\Omega}
      - \ii\mathcal{L}_{\tilde{v}} \bar{\Omega}\wedge\sigma\wedge\Omega
      -  4\imath_{\tilde{v}} F\wedge\sigma
      - \tfrac{1}{3}\lambda_{3}\int_M \imath_{\tilde{v}}\sigma \vol_5 \\
 & = \int_{M}\imath_{\tilde{v}} \sigma\bigl(
      \tfrac{1}{4}\ii \dd\Omega\wedge\bar{\Omega} 
     - \tfrac{1}{2}F -\tfrac{1}{3}\lambda_3  \vol_5\bigr),
\end{split}
\end{equation}
where we have simplified using the previous conditions. Requiring that the expression above vanishes for all $\tilde{v}$ fixes the flux to $F=\tfrac{1}{2}\ii \, \dd\Omega\wedge\bar{\Omega}-\tfrac{2}{3}\lambda_3 \vol_5$.

For the vector-multiplet structure~\eqref{eq:VM_diff}, using the expression for the twisted Dorfman derivative, we find
\begin{equation}
\hat{\Dorf}_{\tilde{K}}\tilde{K}=\mathcal{L}_{\xi}\xi+\mathcal{L}_{\xi}(-\sigma\wedge\omega)-\imath_{\xi}\bigl(\dd(-\sigma\wedge\omega)-\imath_{\xi}F_{5}\bigr) = -\dd\omega,
\end{equation}
which vanishes if $\omega$ is closed. Finally, the  condition~\eqref{eq:HV_diff_5} on $\Dorf_KJ_\alpha$, combined with the conditions from the hyper- and vector-multiplet structures, fixes the remaining $\SU2$ torsion classes and the five-form flux in terms of the cosmological constant. Setting $\lambda_3=3m$, we have
\begin{equation}
\dd\sigma=2m \omega , \qquad \dd \Omega = 3\ii m \sigma\wedge\Omega , \qquad F = 4m \vol_5.
\end{equation}
We see that we reproduce the full set of Sasaki--Einstein integrability conditions~\eqref{eq:SE5-integ}. 

In summary, we have shown that a background consisting of a five-dimensional manifold with an $\SU2$ structure, and generic five-form flux defines a generalised $\USp6$ \HV{} structure. Furthermore, requiring that the \HV{} structure is ESE implies that the $\SU2$ structure is Sasaki--Einstein and the five-form flux takes the correct supersymmetric value.

\subsection{Generic fluxes in type IIB}

Although we will not give the full analysis, let us makes some comments on the case of generic fluxes in type IIB, first considered in~\cite{GMSW06}. In this case, the Killing spinors defines a local $\Uni{1}$ structure and there are a large number of tensors that can be defined in terms of spinor bilinears. The \HM{} and \VM{} structures for generic backgrounds, as in the Sasaki--Einstein case, can again be written in terms of appropriate spinor bilinears. In particular, it is relatively easy to show that the untwisted \VM{} structure takes the form 
\begin{equation}
   \tilde{K} = \xi + \ee^{2\Delta'}\lambda^i + \ee^{4\Delta'}\rho ,
\end{equation}
where, in terms of the fermion bilinears, using the notation of~\cite{GMSW06}, we have\footnote{Note that $\Delta'=\Delta-\frac{1}{4}\phi$ is the warp factor in the Einstein frame, corresponding to that used in~\cite{GMSW06}.}
\begin{equation}
   \xi = K_5^\sharp, \qquad
   \lambda^1 = \re K_3, \qquad
   \lambda^2 = \im K_3, \qquad
   \rho = -\star V, \qquad,
\end{equation}
where $\xi$ is again the Killing vector for the R-symmetry. As we pointed out in \eqref{eq:Lie-xi}, the fact that $K$ is a generalised Killing vector means that the generalised Lie derivative along $\tilde{K}$ reduces to a conventional Lie derivative along the Killing direction. For this to be true, the tensor $\tilde{R}$, defined in \eqref{eq:tildeR}, must vanish.\footnote{Note that one should include the axion-dilaton in $\tilde{K}$ by an appropriate $\sln2$ transformation, as it is not included in $\tilde{R}$.} This follows from the differential conditions
\begin{align}
\dd(\ee^{2\Delta'} K_3) &= \ii Q\wedge K_{3}-P\wedge\bar{K}_3-\ii_{K_5}G ,\\
\dd(\ee^{4\Delta'}\star V) &= -\imath_{\xi}F + \tfrac{\ii}{2}\ee^{2\Delta}(G\wedge\bar{K}_3 - \bar{G}\wedge K_3),
\end{align}
where $G$ is the complex three-form flux and the other forms are defined in \cite{GMSW06}. These conditions are most easily derived directly from the Killing spinor equations.

Recall that there is also a complex bilinear two-form $W$ satisfying
\begin{equation}
   D(\ee^{6\Delta'} W) + P \wedge \ee^{6\Delta'} \bar{W} = (f/4m) G ,  
\end{equation}
where $f$ is a constant related to the five-form flux on $M$. This condition implies that $B^1+\ii B^2=(4m/f)\ee^{6\Delta'} W$ are potentials for the three-form flux $G$~\cite{GGPSW10}.  Using these potentials in~\eqref{eq:IIB-twist}, and the explicit forms of the bilinears given in~\cite{GMSW06}, we then find that the full twisted \VM{} structure is given by\footnote{Note that this includes the dressing by the axion-dilaton degrees of freedom. There is a slight subtlety that here we first twist by the $B^i$ potentials defined by $W$ and then dress by the axion-dilaton, whereas in~\cite{AW15}, the transformations were made in the opposite order. Thus strictly the potentials defined by $W$ differ from those in~\cite{AW15} by the axion-dilaton dressing.}
\begin{equation}
   K = \xi - \sigma\wedge \omega + \imath_\xi C  ,
\end{equation}
where $\dd\sigma=(8m^2/f) \omega$, $C$ is the four-form potential for the five-form flux $F=\dd C - \frac{1}{2}F^i\wedge B^j$. In the notation of~\cite{GMSW06}, $\sigma$ and $\omega$ are defined as
\begin{equation}
   \sigma = \tfrac{4m}{f}\ee^{4\Delta'} K_4 , \qquad
   \omega = - \ee^{4\Delta'} V .
\end{equation}
We see that the form of $K$ is identical to the Sasaki--Einstein case. Furthermore, in~\cite{GGPSW09,GGPSW10}, it was shown that $\sigma$ is a contact structure, even in the case of generic flux, and $\xi$ is the corresponding Reeb vector. The corresponding contact volume is
\begin{equation}
   \tfrac{1}{2}\sigma\wedge\dd\sigma\wedge\dd\sigma 
      = - \left(\tfrac{8m^2}{f}\right)^2\ee^{3\Delta'}\vol_5 
      = - \left(\tfrac{8m^2}{f}\right)^2 c(K) , 
\end{equation}
where $\vol_5$ is the volume of $M$ in the Einstein frame, and we see that it is the $\Ex{6(6)}$-invariant volume up to a constant. 

\subsection{Generic fluxes in M-theory}

We now consider the most general supersymmetric solutions of eleven-dimensional supergravity of the form $\text{AdS}_{5}\times M$, as first discussed in~\cite{GMSW04c}. In this case, the internal six-dimensional space $M$ has a local $\SU{2}$ structure characterised by tensor fields constructed as bilinears of the Killing spinor on $M$. The metric on $M$ always admits a Killing vector corresponding to the R-symmetry of the dual $\mathcal{N}=1$ superconformal field theory. As we will see, in this case, the embedding of the $\SU2$ structure into the \HM{} and \VM{} structures is fairly intricate. 

Let us start by summarising the structure of the solution and the relevant spinor bilinears. The metric is a warped product of the form \eqref{eq:metric_ansatz} with $D=5$. Locally, the internal metric can be written as 
\begin{equation}
   \dd s^2(M) = \dd s^2_{\SU2} + \zeta_1^1 + \zeta_2^2 ,
\end{equation}
where the $\SU2$ structure on $\dd s^2_{\SU2}$ is captured by a complex two-form $\Omega$ and a real fundamental two-form $\omega$. The volume form is given by
\begin{equation}
   \vol_6 = \tfrac{1}{2}\omega\wedge \omega\wedge \zeta_{1}\wedge \zeta_{2}
      = \tfrac{1}{4}\Omega\wedge\bar{\Omega} \wedge \zeta_{1}\wedge \zeta_{2} . 
\end{equation}
We also have an almost complex structure for $\dd s^2_{\SU2}$ given by
\begin{equation}
   I_{\phantom{m}n}^{m} = -\omega_{\phantom{m}n}^{m}
      = \tfrac{1}{4}\ii(\bar{\Omega}^{mp}\Omega_{np}-\Omega^{mp}\bar{\Omega}_{np}).
\end{equation}
The set of spinor bilinears defined in~\cite{GMSW04c} are\footnote{Note that, compared with~\cite{GMSW04c}, we have relabelled $\lambda$ to $\Delta$, $\zeta$ to $\theta$ and $K_i$ to $\zeta_i$. We have also absorbed an overall warp factor into $\dd s^2(M)$.}
\begin{equation}
\begin{aligned}
   \sin\theta & =\bar{\epsilon}^{+}\epsilon^{-} , & &&
   Y = \omega - \sin\theta\, \zeta_{1}\wedge\zeta_{2} 
       &= - \ii\bar{\epsilon}^{+}\gamma_{(2)}\epsilon^{+} , \\ 
   \tilde{\zeta}_{1} = \cos\theta\, \zeta_1 
       &= \bar{\epsilon}^{+}\gamma_{(1)}\epsilon^{+} , & &&
   Y^{\prime} =\zeta_1 \wedge \zeta_2 - \sin\theta\, \omega 
       &= \ii\bar{\epsilon}^{+}\gamma_{(2)}\epsilon^{-} , \\
   \tilde{\zeta}_{2} = \cos\theta\, \zeta_2 
       &= \ii\bar{\epsilon}^{+}\gamma_{(1)}\epsilon^{-} , & &&
   X = - \Omega \wedge (\sin\theta\, \zeta_1-\ii\zeta_2) 
       &= \epsilon^{+\trsp}\gamma_{(3)}\epsilon^{+}, \\
   \tilde{\Omega} = \cos\theta\, \Omega 
       & = \epsilon^{-\trsp}\gamma_{(2)}\epsilon^{+}, & &&
   V = \cos\theta\, \omega \wedge \zeta_2
       &= \bar{\epsilon}^{+}\gamma_{(3)}\epsilon^{-}, 
\end{aligned}
\end{equation}
where $\gamma$ are gamma matrices for $\Cliff{6}$ in an orthonormal frame for $M$ and the Killing spinor on $M$ is split into $\epsilon^{+}$ and $\epsilon^{-}$, where $\epsilon^{-}\propto\gamma_{7}\epsilon^{+}$. In the following we will also need four other, related bilinears
\begin{equation}
\begin{aligned}
   \ii\star X & =\epsilon^{-\trsp}\gamma_{(3)}\epsilon^{+}, & &&
   - \tilde{\zeta}_{1}\wedge Y 
       &= \ii\bar{\epsilon}^{+}\gamma_{(3)}\epsilon^{+}, \\
   \tfrac{1}{3!}Y\wedge Y\wedge Y 
       &= \ii\bar{\epsilon}^{+}\gamma_{(6)}\epsilon^{+}, & &&
   Z = {}\star \tilde{\zeta}_1 
       &= \ii\bar{\epsilon}^{+}\gamma_{(5)}\epsilon^{-}. 
\end{aligned}
\end{equation}

The differential conditions on the $\SU2$ structure derived from the Killing spinor equations are given in (B.9) -- (B.16) of~\cite{GMSW04c}: we reproduce those that we need here\footnote{As mentioned, we have absorbed an overall warp factor into the metric on $M$, so that the powers of $\Delta$ appearing here are different to those in~\cite{GMSW04c}.}
\begin{equation}
\begin{aligned}\label{eq:M_KS1}
\dd(\ee^{3\Delta}\sin\theta) & =2m\ee^{2\Delta}\tilde{\zeta}_{1}, & \dd(\ee^{5\Delta}\tilde{\zeta}_{2}) & =\star F+4m\ee^{4\Delta}Y,\\
\dd(\ee^{3\Delta}X) & =0, & \dd(\ee^{3\Delta}V) & =\ee^{3\Delta}\sin\theta F+2m\ee^{2\Delta}\star Y^{\prime}.
\end{aligned}
\end{equation}
One can use the Killing spinor equations to derive additional identities for forms that were not considered in~\cite{GMSW04c} (but are implied by the conditions therein). We find 
\begin{equation}\label{eq:M_Killing_diff}
   \dd(\ee^{\Delta}Y^{\prime})=-\imath_{\xi}F,\qqq
   \dd(\ee^{\Delta}Z)=\ee^{\Delta}Y^{\prime}\wedge F,
\end{equation}
where $\xi=\ee^{\Delta}\tilde{\zeta}_2^\sharp$ is the Killing vector that preserves the full solution
\begin{equation}
   \mathcal{L}_{\xi}F=\mathcal{L}_{\xi}\Delta=\mathcal{L}_{\xi}g=0,
\end{equation}
and generates the $\Uni1$ R-symmetry. Since the R-symmetry maps $\epsilon^\pm$ to $\ee^{\ii\alpha}\epsilon^\pm$, Lie derivatives of the spinor bilinears vanish except for 
\begin{equation}
\label{eq:xi-bilinear}
   \mathcal{L}_{\xi} \tilde{\Omega} = 3\ii m \tilde{\Omega}, \qquad
   \mathcal{L}_\xi X = 3\ii m X , 
\end{equation}
as can be derived from the conditions in~\cite{GMSW04c}. 

\subsubsection{Embedding as a generalised structure}

The untwisted \HV{} structure is defined in terms of the spinor bilinears as follows. For the $\SUstar{6}$ structure we have
\begin{equation}
\begin{split}
   \tilde{J}_{+} & =\tfrac{1}{2}\kappa\bigl(
      \tilde{\Omega}_R-\ii\star X+\ii\star X^{\sharp}\bigr), \\
   \tilde{J}_{3} & = - \tfrac{1}{2}\kappa Y_R
       + \tfrac{1}{2}\kappa\bigl(\tilde{\zeta}_{1}\wedge Y 
            - \tilde{\zeta}^{\sharp}_{1}\wedge Y^{\sharp}\bigr)
       - \tfrac{1}{2}\kappa\bigl(\tfrac{1}{3!}Y\wedge Y\wedge Y 
            + \tfrac{1}{3!}Y^{\sharp}\wedge Y^{\sharp}\wedge Y^{\sharp}\bigr) , 
\end{split}
\end{equation}
where $\kappa^{2}=\ee^{3\Delta}\vol_{6}$ is the $\Ex{6(6)}$-invariant volume and $\tilde{\Omega}_R$ and $Y_R$ are sections of $TM\otimes T^*M$, constructed by raising the first index of the corresponding two-form with the metric, that is $(\tilde{\Omega}_R)^m_{\phantom{m}n}=g^{mp}\tilde{\Omega}_{pn}$ and $(Y_R)^m_{\phantom{m}n}=g^{mp}Y_{pn}$. The $\Fx{4(4)}$ structure is given by the generalised Reeb vector 
\begin{equation}
\tilde{K}=\xi - \ee^\Delta Y' + \ee^\Delta Z . 
\end{equation}
Using the adjoint action, $\ex{6(6)}$ Killing form and cubic invariant given in~\cite{AW15}, one can check the $J_\alpha$ satisfy an $\su{2}$ algebra and that both structures are correctly normalised. To be sure that together they define an $\USp6$ structure we also need to check the first compatibility condition in~\eqref{eq:compatibility}, or equivalently $\tilde{J}_{+} \cdot \tilde{K}=0$. Splitting into vector, two-form and five-form components, we find
\begin{equation}
\begin{aligned}
   \bigl.\tilde{J}_{+}\cdot \tilde{K}\bigr|_{TM} 
      & \propto\tilde{\Omega}_R\cdot\tilde{\zeta}_2^\sharp
         -\ii(\star X)^{\sharp}\lrcorner Y'=0, \\
   \bigl.\tilde{J}_{+}\cdot \tilde{K}\bigr|_{\mbox{$\wedge$}^2 T^* M} 
      & \propto\tilde{\Omega}_R\cdot Y' 
          + \ii\tilde{\zeta}_2^\sharp\lrcorner(\star X)
          -\ii(\star X)^{\sharp}\lrcorner Z=0,\\
   \bigl.\tilde{J}_{+}\cdot \tilde{K}\bigr|_{\mbox{$\wedge$}^5T^*M} 
      & \propto\tilde{\Omega}_R\cdot Z+\ii(\star X)\wedge Y'=0,
\end{aligned}
\end{equation}
where we have used the expressions for the spinor bilinears in terms of the $\SU2$ structure to see that each term vanishes. The full structures will be twisted by the three-form gauge potential $A$ as in~\eqref{eq:M-twist}. However, it is again actually easier to work with the untwisted structures and use the twisted generalised Lie derivative in the differential conditions. 

\subsubsection{Integrability}

We now turn to the integrability conditions starting with the moment maps~\eqref{eq:mm}. Let $\tilde{V}=\tilde{v}+\tilde{\omega}+\tilde{\sigma}$ be an untwisted generalised vector. The function~\eqref{eq:cosmo_d5} then takes the form
\begin{equation}
   \gamma(\tilde{V}) = -\tfrac{1}{3} \int \ee^{2\Delta}
        \bigl(\tilde{\zeta}_{1}\wedge\tilde{\sigma}
        +\star Y^{\prime}\wedge\tilde{\omega}
        -\imath_{\tilde{v}}Y^{\prime}\wedge Z\bigr).
\end{equation}
We first consider $\mu_{3}$. The moment map is a sum of terms that depend on arbitrary $\tilde{v}$, $\tilde{\omega}$ and $\tilde{\sigma}$, so we can consider each component in turn. The $\tilde{\sigma}$ component is \begin{equation}
\begin{split}
   \mu_3(\tilde{\sigma})-\lambda_3\gamma(\tilde\sigma) 
      &= \tfrac{1}{16}\ii\int_{M} 
         \kappa^2\bigl(
            \star\bar{X}^{\sharp}\wedge \star X^{\sharp}\bigr)
               \lrcorner\dd\tilde{\sigma}
          + \tfrac{1}{3}\lambda_{3}\int_{M}
              \ee^{2\Delta}\tilde{\zeta}_{1}\wedge\tilde{\sigma}\\
      &= \tfrac{1}{2}\int_{M}
          \ee^{3\Delta}\sin\theta\,\dd\tilde{\sigma}
          + \tfrac{1}{3}\lambda_{3}\int_{M}
              \ee^{2\Delta}\tilde{\zeta}_{1}\wedge\tilde{\sigma}\\
      & = - \tfrac{1}{2}\int_{M}
              \dd(\ee^{3\Delta}\sin\theta)\wedge\tilde{\sigma}
           + \tfrac{1}{3}\lambda_{3}\int_{M}
              \ee^{2\Delta}\tilde{\zeta}_{1}\wedge\tilde{\sigma}.
\end{split}
\end{equation}
Remembering that $\lambda_3=3m$, this vanishes for
\begin{equation}
   \dd(\ee^{3\Delta}\sin\theta)=2m\ee^{2\Delta}\tilde{\zeta}_{1}.
\end{equation}
This is the first differential condition in~\eqref{eq:M_KS1}. The $\tilde{\omega}$ component is
\begin{equation}
\begin{split}
   \mu_3(\tilde{\omega})-\lambda_3\gamma(\tilde\omega) 
   &= \tfrac{1}{16}\ii\int_{M}
        \kappa^{2}\Bigl(
           \ii\bigl(\bar{\tilde{\Omega}}_R\cdot \star X^{\sharp}
           + \tilde{\Omega}_R\cdot \star\bar{X}^{\sharp}\bigr)
              \lrcorner\dd\tilde{\omega}
           + \bigl( \star\bar{X}^{\sharp}\wedge \star X^{\sharp}\bigr)
               \lrcorner(\tilde{\omega}\wedge F) \Bigr) \\
   & \qquad  {} + \tfrac{1}{3}\lambda_{3}\int_{M}
        \ee^{2\Delta}\star Y'\wedge\tilde{\omega} \\
   &= -\tfrac{1}{2}\int_{M}\Bigl(
           \ee^{3\Delta}V\wedge\dd\tilde{\omega}
           - \sin\theta\ee^{3\Delta}\tilde{\omega}\wedge F \Bigr)
        + \tfrac{1}{3}\lambda_{3}\int_{M}
           \ee^{2\Delta}\star Y'\wedge\tilde{\omega} \\
   &= -\tfrac{1}{2}\int_{M}\Bigl(
          \dd(\ee^{3\Delta}V)\wedge\tilde{\omega}
          - \sin\theta\ee^{3\Delta}\tilde{\omega}\wedge F \Bigr)
       + \tfrac{1}{3}\lambda_{3}\int_{M}
          \ee^{2\Delta}\star Y'\wedge\tilde{\omega}.
\end{split}
\end{equation}
This vanishes for
\begin{equation}
\dd(\ee^{3\Delta}V)=\ee^{3\Delta}\sin\theta F+2m\ee^{2\Delta}\star Y'.
\end{equation}
This is the fourth differential condition in \eqref{eq:M_KS1}. The $\tilde{v}$ component is rather long but can be shown to vanish as a result of the differential conditions in~\eqref{eq:M_KS1}. For the $\mu_{+}$ moment map, the contribution from terms containing $\tilde{\sigma}$ vanishes without imposing any differential conditions. The contribution from the $\tilde{\omega}$ terms simplifies to
\begin{equation}
   \mu_+(\tilde{\omega})  
      = -\tfrac{\ii}{2}\int_{M}\ee^{3\Delta}X\wedge\dd\tilde{\omega}
      = -\tfrac{\ii}{2}\int_{M}\dd(\ee^{3\Delta}X)\wedge\tilde{\omega}.
\end{equation}
This vanishes after imposing the third differential condition in \eqref{eq:M_KS1}
\begin{equation}
   \dd(\ee^{3\Delta}X)=0.
\end{equation}
The $\tilde{v}$ component is again somewhat involved but can be shown to vanish as a result of the conditions~\eqref{eq:M_KS1}.

For the vector-multiplet structure we first use the condition~\eqref{eq:Lie-xi}, which, substituting for $\tilde{K}$ in~\eqref{eq:tildeR}, gives 
\begin{equation}
   \tilde{R} = - \dd(\ee^{\Delta}Y^{\prime})-\imath_{\xi}F
       + \dd(\ee^{\Delta}Z) - \ee^{\Delta}Y^{\prime}\wedge F = 0, 
\end{equation}
which reproduces the two equations in~\eqref{eq:M_Killing_diff}. We then have 
\begin{equation}
   \hat{\Dorf}_{\tilde{K}} \tilde{K} = \mathcal{L}_\xi \tilde{K} = 0 , 
\end{equation}
since the bilinears $\xi=\ee^\Delta\zeta_2^\sharp$, $Y'$ and $Z$ are all invariant. Finally we have the condition~\eqref{eq:HV_diff_5} which, given~\eqref{eq:xi-bilinear}, reads 
\begin{equation}
\hat{\Dorf}_{K}\tilde{J}_{+}=\mathcal{L}_{\xi}\tilde{J}_{+}=3\ii m \tilde{J}_{+},
\end{equation}
in agreement with $\lambda_{3}=3m$. 

In summary, we have shown that the most general $\AdS5$ solutions of eleven-dimensional supergravity do indeed define an exceptional Sasaki--Einstein space. 

\section{\texorpdfstring{$\AdS4$}{AdS4} backgrounds as ESE spaces}\label{sec:AdS4}

We now discuss the structure of exceptional Sasaki--Einstein (ESE) backgrounds for $\AdS4$. We first show how the standard M-theory Sasaki--Einstein reduction with seven-form flux embeds as an ESE background, and comment on how this extends to the generic case, given in~\cite{GMPS14}.

\subsection{Sasaki--Einstein in M-theory}

We now briefly discuss the structure of exceptional Sasaki--Einstein (ESE) backgrounds for $\AdS4$, focussing on the example of conventional Sasaki--Einstein geometry in M-theory. These are supersymmetric solutions preserving at least eight supercharges~\cite{AFHS99}, and are dual to a three-dimensional superconformal field theory living on a stack of M2-branes placed at the tip of the corresponding Calabi--Yau cone. 

The metric is a product of the form \eqref{eq:metric_ansatz} with $D=4$ and a constant warp factor, which we take to be zero. Seven-dimensional Sasaki--Einstein spaces admit a nowhere-vanishing vector field $\xi$, known as the Reeb vector, a complex three-form $\Omega$ and real two-form $\omega$, which together define an $\SU3\subset\GL{7;\bbR}$ structure. They satisfy the algebraic conditions 
\begin{equation}\label{eq:SE7_alg}
   \tfrac{1}{8}\ii\Omega\wedge\bar{\Omega} 
      = \tfrac{1}{3!} \omega\wedge\omega\wedge\omega , \qquad
   \imath_{\xi}\Omega=\imath_{\xi}\omega=0, \qquad
   \imath_{\xi}\sigma=1,
\end{equation}
where $\sigma$ is the one-form constructed from $\xi$ by lowering the index with the metric. In addition one has the differential conditions 
\begin{equation}
\label{eq:SE7-integ}
\dd\sigma=m\omega , \qquad \dd\Omega=2\ii m\sigma\wedge\Omega, 
\end{equation}
where $m$ is the inverse $\AdS4$ radius, usually normalised to $m=2$. Such a compactification is supersymmetric provided there is a seven-form flux given by 
\begin{equation}
   \dd \tilde{A}=\tilde{F}=-3m\vol_{7},
\end{equation}
where $\vol_{7}=\tfrac{1}{3!}\sigma\wedge\omega\wedge\omega\wedge\omega$. (Recall that  $\tilde{F}$ is the Hodge-dual of the four-form flux $\mathcal{F}=6m\vol(\text{AdS}_{4})$ in eleven-dimensions.) These conditions imply that the Reeb vector $\xi$ is a Killing vector that preserves $\sigma$ and $\omega$, but rotates $\Omega$ by a phase
\begin{equation}
\mathcal{L}_{\xi}\sigma=\mathcal{L}_{\xi}\omega=\mathcal{L}_{\xi}g=0,\qquad
\mathcal{L}_{\xi}\Omega=2\ii m \Omega.
\end{equation}
The rotation of $\Omega$ corresponds to the R-symmetry of the $\mathcal{N}=2$ solution. In what follows we also need the (transverse) complex structure
\begin{equation}
I_{\phantom{m}n}^{m}=-\omega_{\phantom{m}n}^{m}
    =\tfrac{\ii}{4}(\bar{\Omega}{}^{mp}\Omega_{np}-\Omega^{mp}\bar{\Omega}_{np}),
\end{equation}
which satisfies $I_{\phantom{p}m}^{p}\Omega_{pn}=\ii\Omega_{mn}$. For simplicity of presentation, we assume that the four-form flux and warp factor vanish, though one can show that these also follow from the integrability conditions. 

The \HV{} structure defined by the $\SU3$ structure is actually the same as an example considered in~\cite{AW15}, namely a Calabi--Yau threefold times a circle. The difference between the two is in the differential conditions on the $\SU3$ invariant forms. We have the untwisted tensors 
\begin{equation}
\begin{split}
\tilde{J}_{+} & =\tfrac{\kappa}{2}\Omega-\tfrac{\kappa}{2}\Omega^{\sharp},\\
\tilde{J}_{3} & =\tfrac{\kappa}{2}I-\tfrac{\kappa}{2}\tfrac{\ii}{8}\Omega\wedge\bar{\Omega}-\tfrac{\kappa}{2}\tfrac{\ii}{8}\Omega^{\sharp}\wedge\bar{\Omega}^{\sharp},
\end{split}
\end{equation}
where $\kappa^{2}=\vol_{7}$ is the $\Ex{7(7)}$-invariant volume and 
\begin{equation}
\label{eq:X-SE7}
\tilde{X}=\xi+\ii\omega-\tfrac{1}{2}\sigma\wedge\omega\wedge\omega-\ii\sigma\otimes\vol_{7}.
\end{equation}
Using the adjoint action, the symplectic invariant and  $\ex{7(7)}$ Killing form given in~\cite{AW15}, one can check that $\tilde{J}_\alpha$ generate an $\su2$ algebra and that both structures are correctly normalised and are compatible, as in~\eqref{eq:J_conditions} and~\eqref{eq:compatibility}. 

We now show how the integrability conditions on the $\SU3$ structure arise by requiring $\{J_\alpha,K\}$ to be ESE. Starting with the moment maps~\eqref{eq:mm}, we note that if $\tilde{V}=\tilde{v}+\tilde{\omega}+\tilde{\sigma}+\tilde{\tau}$ is an arbitrary untwisted generalised vector, then 
\begin{equation}
   \gamma(\tilde{V}) = \int_M s(\tilde{V},\tilde{\hat{K}})
      = -\tfrac{1}{4}\int_M
         (\imath_{\tilde{v}}\sigma\vol_{7}+\tilde{\sigma}\wedge\omega).
\end{equation}
Starting with $\mu_{3}$, the terms that depend on $\tilde{\sigma}$ are
\begin{equation}
\begin{split}
   \mu_{3}(\tilde{\sigma})-\lambda_3\gamma(\tilde{\sigma}) 
      & = - \tfrac{1}{16}\ii\int_{M}
          \kappa^{2}(\bar{\Omega}^{\sharp}\wedge\Omega^{\sharp})
             \lrcorner\dd\tilde{\sigma})
          + \tfrac{1}{4}\lambda_{3}\int_{M}\tilde{\sigma}\wedge\omega \\
      &= - \tfrac{1}{2}\int_{M} \dd \tilde{\sigma}\wedge\sigma
           + \tfrac{1}{4}\lambda_{3}\int_{M}\tilde{\sigma}\wedge\omega\\
      &= -\tfrac{1}{2}\int_{M}\tilde{\sigma}\wedge \dd\sigma
           + \tfrac{1}{4}\lambda_{3}\int_{M}\tilde{\sigma}\wedge\omega,
\end{split}
\end{equation}
which vanishes for $\dd\sigma=\tfrac{1}{2}\lambda_3 \omega$. The $\mu_{+}$ moment map is (using the notation of~\cite{AW15})
\begin{equation}
\begin{split}
    \mu_{+}(\tilde{V}) 
        &=- \tfrac{1}{2}\ii\int_{M} 
           - \tfrac{1}{4}\kappa^{2}\tr\bigl( 
               I\cdot(j\Omega^{\sharp}\lrcorner j\dd\tilde{\omega})\bigr)
               +\tfrac{1}{2^{2}\cdot3!}\kappa^{2}
                  (\omega^{\sharp}\wedge\omega^{\sharp}\wedge\omega^{\sharp})
                  \lrcorner(\dd\tilde{\omega}\wedge\Omega)\\
        &= - \tfrac{1}{8}\ii\int_{M}3\ii\kappa^{2}\Omega^{\sharp}
               \lrcorner\dd\tilde{\omega}
           + \sigma\wedge\dd\tilde{\omega}\wedge\Omega\\
        &= \tfrac{1}{2}\ii\int_{M}\sigma\wedge\Omega\wedge\dd\tilde{\omega},
\end{split}
\end{equation}
which, using $\dd\sigma\propto\omega$ from above, vanishes for $\sigma\wedge\dd\Omega=0$. In the language of~\cite{BCL06}, this fixes the torsion classes $\{\mathcal{W}_1,\mathcal{W}_2,\mathcal{W}_5\}$ to zero. Finally, the $\tilde{v}$ components of $\mu_3$ are
\begin{equation}
\begin{split}
   \mu_{3}(\tilde{v})-\lambda_3\gamma(\tilde{v}) 
      &= -\tfrac{1}{16}\ii\int_{M}
            \kappa\bar{\Omega}^{\sharp}\lrcorner
               \mathcal{L}_{\tilde{v}}(\kappa\Omega)
            + \mathcal{L}_{\tilde{v}}(\kappa\Omega^{\sharp})
               \lrcorner\kappa\bar{\Omega}
               -\kappa^{2}(\bar{\Omega}^{\sharp}\wedge\Omega^{\sharp})
                  \lrcorner\imath_{\tilde{v}}\tilde{F}\\
      & \qquad -\tfrac{1}{4}\lambda_{3}\int_{M}\imath_{\tilde{v}}\sigma\vol_{7} \\
      &= \tfrac{1}{8}\int_{M}\bigl( 
         \imath_{\tilde{v}} \sigma \, \dd \Omega \wedge \bar{\Omega}  
         + 4 \, \imath_{\tilde{v}}\sigma \, \tilde{F} \bigr)
         - \tfrac{1}{4}\lambda_{3}\int_{M}\imath_{\tilde{v}}\sigma\vol_{7},
\end{split}
\end{equation}
where we have used the previous results to reach the final line. Requiring this to vanish fixes the flux to $\tilde{F} = \tfrac{1}{2} \lambda_3 \vol_7 - \tfrac{1}{4}\dd\Omega\wedge\bar{\Omega}$.

For the vector-multiplet structure, using the expression for the twisted generalised Lie derivative~\eqref{eq:twisted-Dorf} and~\eqref{eq:tildeR}, we find 
\begin{equation}
   \hat{\Dorf}_{\tilde{K}}\tilde{K} 
      = \mathcal{L}_\xi \xi 
         + \mathcal{L}_\xi(-\tfrac{1}{2}\sigma\wedge\omega\wedge\omega)
         - \imath_\xi \bigl( \dd(-\tfrac{1}{2}\sigma\wedge\omega\wedge\omega)
            - \imath_\xi \tilde{F} \bigr)
      = - \dd\omega\wedge\omega ,
\end{equation} 
so that integrability implies $\dd\omega\wedge\omega=0$. In the language of~\cite{BCL06}, the torsion classes corresponding to $\{\mathcal{W}_4,E+\bar{E},V_2,T_2\}$ must vanish. Finally, the conditions from~\eqref{eq:HV_diff_4} combined with those from the \HM{} and \VM{} structures fix the remaining $\SU3$ torsion classes to $S=0$ and $E=\ii\lambda_3$, so that, with $\lambda_3=2m$, we have 
\begin{equation}
\dd\sigma=m \omega , \qquad \dd \Omega = 2\ii m \sigma\wedge\Omega , \qquad \tilde{F} = -3m \vol_7.
\end{equation}
We see we reproduce the full set of Sasaki--Einstein integrability conditions. 

In summary, we have shown that a background consisting of a seven-dimensional manifold with an $\SU3$ structure and generic seven-form flux defines a generalised $\SU6$ structure. Furthermore, requiring that the \HV{} structure is ESE implies the manifold must be Sasaki--Einstein and the seven-form flux matches that of the standard supersymmetry-preserving solution.

\subsection{Generic fluxes in M-theory}

Although we will not give the full analysis, let us now discuss some aspects of how the previous analysis extends to the case of generic fluxes in M-theory, first considered in~\cite{GMPS14}. In this case, the Kiling spinors define a local $\SU2$ structure. The \HM{} and \VM{} structures for generic backgrounds, as in the Sasaki--Einstein case, can be written in terms of appropriate spinor bilinears. Assuming the seven-form $\tilde{F}$ is non-zero, it is relatively straightforward to show that the complex untwisted \VM{} structure takes the form 
\begin{equation}
   \tilde{X} = \xi + \ee^{3\Delta}Y + \ee^{6\Delta}Z - \ii\ee^{9\Delta}\tau ,
\end{equation}
where, in terms of the fermion bilinears, using the notation of~\cite{GMPS14}, we have
\begin{equation}
    \xi = \ii \bar{\chi}^c_+ \gamma^{(1)} \chi_-, \qquad
    Y = \ii\bar{\chi}^c_+ \gamma_{(2)} \chi_-, \qquad
    Z = \star Y , \qquad 
    \tau = \xi^\flat \otimes \vol_7 . 
\end{equation}
The tensors $Y$ and $Z$ are generically complex, but, as shown in~\cite{GMPS14}, $\xi$ is real, so there is no vector component in the imaginary part of $X$, consistent with the general argument given at the end of section~\ref{sec:ESE}. The generalised Lie derivative along the real part of $\tilde{X}$ generates the R-symmetry, and so must reduce to a conventional Lie derivative along $\xi$. We indeed find that the tensor $\tilde{R}$, defined in~\eqref{eq:tildeR}, vanishes due to
\begin{align}
\dd (\ee^{3\Delta} Y) &= \imath_\xi F,\\
\dd (\ee^{6\Delta}Z) &= \imath_\xi \tilde{F} -\ee^{3\Delta}Y\wedge F,
\end{align}
where the first is given in~\cite{GMPS14} and the second can be derived from the Killing spinor equations.

Recall also that there is also a spinor bilinear three-form satisfying
\begin{equation}
   \dd \bigl(\ee^{6\Delta}\im(\bar{\chi}_+^c\gamma_{(3)}\chi_-)\bigr) 
      = \tfrac{\tilde{f}}{3m} F . 
\end{equation}
Compared with the expression given in~\cite{GMPS14}, we have reinstated the inverse AdS radius $m$ (set to $m=2$ in~\cite{GMPS14}), and $\tilde{f}$ (denoted by $m$ in~\cite{GMPS14}) parameterises the seven-form flux, namely $\tilde{F}=-\tilde{f}\vol_7$. We see $\frac{3m}{\tilde{f}}\ee^{6\Delta}\im(\bar{\chi}_+^c\gamma_{(3)}\chi_-)$ is a potential for the four-form flux $F$.  Using this potential in~\eqref{eq:M-twist} and the explicit forms of the bilinears given in~\cite{GMPS14}, we then find that the full twisted \VM{} structure is given by
\begin{equation}
   X = \ee^{\tilde{A}}\Bigl[
      \xi + \ii\omega 
      - \tfrac{1}{2}\sigma\wedge\omega\wedge\omega
      - \ii\sigma\otimes 
        \left(\tfrac{1}{3!}\sigma\wedge\omega\wedge\omega\wedge\omega \right)
   \Bigr] , 
\end{equation}
where $\dd \sigma = (3m^2/\tilde{f})\omega$. In particular, the real part is given by 
\begin{equation}
   K = \xi - \tfrac{1}{2} \sigma\wedge\omega\wedge\omega 
       + \imath_\xi \tilde{A} .
\end{equation}
We see that the form of $X$ matches that of the Sasaki--Einstein case~\eqref{eq:X-SE7}. It was shown in~\cite{GMPS14} that $\sigma$ is a contact structure, even in the case of generic flux, and $\xi$ is the corresponding Reeb vector. The corresponding contact volume is
\begin{equation}
   \tfrac{1}{3!}\sigma\wedge\dd\sigma\wedge\dd\sigma\wedge\dd\sigma 
      = \left(\tfrac{3m^2}{\tilde{f}}\right)^3\ee^{9\Delta}\vol_7
	= \left(\tfrac{3m^2}{\tilde{f}}\right)^3 2\sqrt{q(K)} ,
\end{equation}
where $\vol_7$ is the volume of $M$. Again it is simply a constant times the  $\Ex{7(7)}$-invariant volume. 

\section{Central charges, BPS wrapped branes and volume minimisation}\label{sec:K-physics}

Of the many field theory properties that can be determined from the dual geometry, two of the most studied are the central charge $a$ or free energy $\mathcal{F}$ of the theory and the conformal dimension of operators that arise from supersymmetric wrapped branes. The key point of this section is that they are all encoded, in a universal way, by the generalised Reeb vector $K$. This also leads to a conjecture as to how the dual description of $a$-maximisation in $D=4$ and $\mathcal{F}$-maximisation in $D=3$ appears. 

We have considered three ESE geometries in this paper: $\AdS5$ in type IIB and M-theory and $\AdS4$ in M-theory. The generic generalised Reeb vector in each case is given by 
\begin{equation}\label{gen-reeb}
   K = \begin{cases}
      \xi - \sigma\wedge \omega + \imath_\xi C  , 
      & \text{$\AdS5$ in type IIB} ,\\
      \xi - \ee^\Delta Y' + \ee^\Delta Z + \imath_\xi A 
        - A\wedge \ee^\Delta Y' , 
      & \text{$\AdS5$ in M-theory} ,\\
      \xi - \tfrac{1}{2}\sigma\wedge\omega\wedge\omega 
       + \imath_\xi\tilde{A} , 
      & \text{$\AdS4$ in M-theory} , 
   \end{cases}
\end{equation}
where in the last case we are assuming the seven-form flux $\tilde{F}$ is non-trivial and in the first that five-form flux $F$ is non-trivial. Each $K$ is a generalised Killing vector that generates the global R-symmetry of the dual field theory. It is a combination of diffeomorphism (parameterised by $\xi$) and gauge transformation (parameterised by the $p$-form components), under which the transformations of the metric $g$ and gauge potentials vanish, as in~\eqref{eq:gen-Killing-m} and~\eqref{eq:gen-Killing-IIB}. For $\AdS5$ in IIB~\cite{GGPSW09,GGPSW10} and $\AdS4$ in M-theory~\cite{GMPS14}, the generic geometry admits a canonical contact structure $\sigma$. As we have already noted, it is striking that this structure is equivalent to specifying the generalised Reeb vector $K$, where the integrability arises from requiring that $K$ is generalised Killing. 

For $\AdS5$ solutions the central charge $a$ of the dual field theory is given by~\cite{HS98}
\begin{equation}
   a = \frac{\pi}{8m^3 G_5} ,
\end{equation}
where $G_5$ is the effective five-dimensional Newton's constant. Using the results of~\cite{GGPSW09} and~\cite{GOV06}, one finds that for both the generic type IIB and M-theory background the inverse of $G_5$ is given by the integral of the $\Ex{6(6)}$-invariant volume 
\begin{equation}
   G^{-1}_{5} 
      \propto \int_M \ee^{3\Delta} \vol
      = \int_M c(K) . 
\end{equation}
As reviewed in appendix~\ref{app:cc}, quantising so we have $N$ units of background flux and fixing this integer $N$ in the expression for $a$ reverses the dependence on the invariant volume. This leads to a universal expression for the central charge in terms of the generalised Reeb vector, applicable to both type IIB and M-theory 
\begin{equation}
\label{eq:a-c}
   a^{-1} \propto \int_M c(K) ,
\end{equation}
where in type IIB the constant of proportionality scales as $N^{-2}$ and in M-theory as $N^{-3}$. Recall that for type IIB, $c(K)$ is proportional to the contact volume $\frac{1}{2}\sigma\wedge\dd\sigma\wedge\dd\sigma$. 

A similar formula for the free energy of the field theory on a three-sphere can be derived for generic $\AdS4$ backgrounds following~\cite{GMPS14}. The real part of the free energy is equal to the gravitational free energy and is given by
\begin{equation}
   \mathcal{F} = \frac{\pi}{2m^2G_4} ,
\end{equation}
where the four-dimensional Newton's constant is given by the $\Ex{7(7)}$-invariant volume 
\begin{equation}
   G^{-1}_{4} 
      \propto \int_M \ee^{2\Delta} \vol_7
      = \int_M 2\sqrt{q(K)} . 
\end{equation}
Fixing the quantised background flux then gives, as in~\cite{JKPS11},
\begin{equation}
\label{eq:F-q}
   \mathcal{F}^{-2} \propto \int_M \sqrt{q(K)} ,
\end{equation}
where the constant of proportionality scales as $N^{-3}$. Again, $\sqrt{q(K)}$ is proportional to the contact volume, $\frac{1}{3!}\sigma\wedge\dd\sigma\wedge\dd\sigma\wedge\dd\sigma$.  Although we have not considered type IIB $\AdS4$ backgrounds, we expect that the same formula for the free energy holds since $q(K)$ (and $c(K)$ in the $\AdS5$ case) are U-duality invariants.

Let us now discuss how the properties of chiral operators in the dual SCFT coming from wrapped branes are encoded by $K$. For definiteness, we will focus on $\AdS5$ in type IIB. A probe D3-brane wrapping a supersymmetric three-cycle $\Sigma_3$ in $M_5$ gives rise to a BPS particle in $\AdS5$. The particle appears as the excitation of a field that couples to a chiral primary operator $\mathcal{O}_3$, and thus the probe D3-brane corresponds to a BPS operator in the dual field theory. The (warped) volume of the wrapped D3-brane is then associated to the conformal dimension of the operator $\Delta(\mathcal{O}_3)$, which in turn is proportional to the R-charge. In order for the three-cycle to be supersymmetric, it must be calibrated by a (generalised) three-form calibration. There are many ways to find this calibration, including using spinor bilinears of the full ten-dimensional Killing spinors or checking the $\kappa$-symmetry conditions directly. 

A similar story applies to probe M2-branes wrapping supersymmetric two-cycles in $M_6$ and probe M5-branes wrapping supersymmetric five-cycles in $M_7$, corresponding to chiral primary operators in the dual four- and three-dimensional SCFTs. For all three cases, the relevant calibration form is known and the conformal dimensions of the corresponding operators are given by
\begin{equation}
\begin{aligned}
\label{eq:cal}
   \text{D3-branes in $\AdS5$~\cite{GGPSW09,GGPSW10}:} && &&
   \Delta(\mathcal{O}_3) 
      &= -  \frac{\tau_\text{D3}}{m} \int_{\Sigma_3} \sigma\wedge\omega , \\
   \text{M2-branes in $\AdS5$~\cite{GOV06}:} && &&
   \Delta(\mathcal{O}_2) 
      &=  \frac{\tau_\text{M2}}{m}\int_{\Sigma_2} \ee^\Delta Y' , \\
   \text{M5-branes in $\AdS4$~\cite{GMPS14}:} && &&
   \Delta(\mathcal{O}_5) 
      &= - \frac{\tau_\text{M5}}{m} \int_{\Sigma_5} \tfrac{1}{2}\sigma\wedge\omega\wedge\omega , 
\end{aligned}
\end{equation}
where $\tau$ is the tension of the brane wrapping the cycle. From \eqref{gen-reeb} we see that the relevant calibration form appears in the generalised Reeb vector $K$, implying that the components of $K$ are the (generalised) calibrations that define supersymmetric cycles. This is not  surprising since $K$ is defined as a bilinear of the Killing spinors and imposing that $\Dorf_K$ reduces to $\mathcal{L}_\xi$ requires the components of $K$ to satisfy equations that resemble generalised calibration conditions. For backgrounds with non-trivial fluxes, the calibration condition is equivalent to asking that the energy of the wrapped brane is minimised. This suggests that the generalised calibration should be given by the twisted $K$.  Notice however that, for the branes we discussed above, most of the potentials have vanishing  
pull-back on the wrapped cycle and hence do not contribute to the conditions \eqref{eq:cal}.  We leave for future work  a more detailed analysis of  how calibrations appear in this language.

As we have seen, the generalised Reeb vector $K$ encodes the central charge or free energy of the dual field theory. For some time, a classic problem in four-dimensional $\mathcal{N}=1$ SCFTs was to find the correct $\Uni1$ symmetry that gives the R-symmetry as the theory flows from the UV to the IR. A general procedure for determining this was given by Intriligator and Wecht~\cite{IW03}, namely $a$-maximisation. For three-dimensional $\mathcal{N}=2$ theories the analogous procedure consists of maximising the free energy~\cite{Jafferis10,JKPS11}. (Both cases can also be thought of as minimising the coefficient $\tau_{RR}$ of the two-point function of the R-symmetry current~\cite{BGISW05}.) The bulk version of this process is known as volume minimisation~\cite{MSY06,MSY08}, and was originally derived for Sasaki--Einstein backgrounds, but a version also appears to hold for the case of generic type IIB backgrounds~\cite{GS12}. The idea is to relax the supersymmetric conditions slightly and show that the resulting supergravity action depends only on the choice of Reeb vector, $\xi$. The actual supersymmetric background then appears after minimising over the possible choices of $\xi$.

This leads to a natural question: what is the dual of $a$-maximisation (or $\mathcal{F}$-maximisation) in our language? Comparing with~\cite{MSY06,MSY08,GS12} there is a very natural candidate for relaxing the supersymmetry conditions, namely simply to drop the normalisation conditions $\kappa^2=c(K)$ in $D=5$ and $\kappa^2=2\sqrt{q(K)}$ in $D=4$, defining a notion of an ``exceptional Sasaki structure''. Following the analogous analysis to that given in appendix~\ref{app:lemma}, we find this requires that the moment map conditions are slightly modified, giving 
\begin{defn}
An \emph{exceptional Sasaki structure} is a pair $\{J_\alpha,K\}$ of H and V structures satisfying $J_\alpha\cdot K=0$ and the integrability conditions 
\begin{align}
   \mu_\alpha(V) &= \lambda_\alpha \int_M \phi(V) \qquad \forall V\in\Gamma(E) , 
   \label{eq:mm-gen-b} \\
   \Dorf_K K &= 0 , 
   \label{eq:VM_diff-b} \\
   \Dorf_K J_\alpha 
      &= \epsilon_{\alpha\beta\gamma} \lambda_\beta J_\gamma, 
      \qqq \Dorf_{\hat{K}}J_\alpha=0,
   \label{eq:HV_diff-b}
\end{align}
where $\phi(V)$ is given by
\begin{equation}
   \phi(V) = \begin{cases}
         \kappa^2q(V,K,K,K)/q(K) ,  & \text{for $D=4$} \\
         \kappa^2c(V,K,K)/c(K) ,  & \text{for $D=5$} 
      \end{cases}
\end{equation}
where $\tr(J_\alpha J_\beta)=-\kappa^2\delta_{\alpha\beta}$ and $\lambda_\alpha$ are real constants, as in the definition of an \ESE{} structure. The condition $\Dorf_{\hat{K}}J_\alpha=0$ is relevant only for $D=4$.
\end{defn}
\noindent
An interesting open question is whether in the $D=5$ type IIB case this agrees with the notion of a generalised Sasaki structure defined in~\cite{GS12}. 
The natural conjecture is then that, over the space of such structures, the supergravity action restricted to the internal space $M$ is given by
\begin{equation}
\label{eq:sugra}
\begin{aligned}
   S_{\text{sugra}} &\propto \int_M \sqrt{q(K)} , 
      & && && \text{and}&& && && 
   S_{\text{sugra}} &\propto \int_M c(K) ,
\end{aligned}
\end{equation}
for $D=4$ and $D=5$ respectively, and so depends only on the generalised Reeb vector. Extremising over the space of $K$ then selects the generalised Reeb vector that corresponds to the actual R-symmetry. 

Motivation for this formulation comes from the fact, already noted in section~\ref{sec:sugra-int}, that the supersymmetry conditions for an \ESE\ structure can be interpreted in terms of gauged $D=4$ or $D=5$ supergravity with infinite dimensional spaces of hyper- and vector-multiplets. Various authors have considered the dual of $a$- and $\mathcal{F}$-maximisation from the point of view of a conventional dual gauged $D=5$ or $D=4$ supergravity~\cite{Tachikawa06,BGIW06,AG15}, and showed explicitly that they correspond to extremising over the space of possible R-symmetries either, in $D=5$, the cubic function that determines the real special geometry of the vector multiplets~\cite{Tachikawa06,BGIW06}, or, in $D=4$, the real function that determines the special K\"ahler geometry of the vector multiplets~\cite{AG15}. In our language, this corresponds to varying $K$ and extremising the integral of either $c(K)$ or $\sqrt{q(K)}$, exactly as we conjecture above. 

Showing that such a procedure works would provide the dual of $a$- and $\mathcal{F}$-maximisation not only for an arbitrary flux background, generalising the Sasaki--Einstein cases in IIB on $\AdS5$ and M-theory on $\AdS7$, but also for the generic M-theory $\AdS5$ background for which no notion of volume minimisation exists. It may also provide insight into exactly what space of solutions one is extremising over in the flux case.

\section{Discussion}\label{sec:discuss}

In this paper we have given a new geometrical interpretation of generic AdS flux backgrounds preserving eight supercharges within generalised geometry. These ``exceptional Sasaki--Einstein'' (ESE) geometries are the natural string generalisations of Sasaki--Einstein spaces in five and seven dimensions. They always admit a ``generalised Reeb vector'' that generates an isometry of the background corresponding to the R-symmetry of the dual field theory. In the language of~\cite{CS15}, ESE spaces are weak generalised holonomy spaces, and the cone over such a space has generalised special holonomy. We have included a number of examples of ESE spaces including conventional Sasaki--Einstein in five and seven dimensions, as well as the most general $\AdS{5}$ solutions in M-theory. We also discussed the structure of the moduli spaces of ESE spaces, pointing out an interesting connection to the ``HK/QK correspondence''~\cite{Haydys08}.

A key application of this analysis is, of course, to the AdS/CFT correspondence and we made some steps in this direction. A particular advantage of the formalism is that the generalised H and V structures defining the background are associated to hypermultiplet and vector-multiplet degrees of freedom in the corresponding gauged supergravity. This provides a natural translation between bulk and boundary properties. We showed for example that the V structure, which is defined by the generalised Reeb vector $K$, encodes the contact structure that appears in generic $D=5$ IIB and $D=4$ M-theory backgrounds~\cite{GGPSW09,GGPSW10,GMPS14}. Furthermore $K$ determines the central charge in $D=5$ and free energy in $D=4$ of the dual theory, and is a calibration for BPS wrapped branes giving the dimension of the dual operators. In the examples with contact structures, this framework allows one to calculate properties of the field theory using the relation between the contact volume and the choice of Reeb vector~\cite{GGPSW09,GGPSW10,GMPS14}. It would be particularly interesting to see if one can extend these techniques to the case of $D=5$ M-theory backgrounds using the generalised Reeb vector. The special role of $K$ also led us, following~\cite{GS12}, to a conjecture for generic form volume minimisation~\cite{MSY06,MSY08}. We hope to come back to this point in the future. In particular, it should be possible to use generalised intrinsic torsion to show that the supergravity actions are given by the integral of the $\Ex{d(d)}$-invariant volume, as in~\eqref{eq:sugra}. 

There are many other directions for future study. An obvious extension is to consider backgrounds with different amounts of supersymmetry, which will be described by new geometric structures 
within generalised geometry. Another is to consider the reduction of generalised structures. Recall that $K$ is always a generalised Killing vector and that the cone over an ESE space has generalised 
special holonomy. In the conventional Sasaki--Einstein case one can use the Reeb vector to define a symplectic reduction of the Calabi--Yau cone. Locally, this gives a four-dimensional geometry that is K\"ahler--Einstein. When one moves to generalised complex geometry and generic flux solutions, there is an analogous result using the theory of generalised quotients that the transverse space admits a generalised Hermitian structure~\cite{GGPSW10}. It would be interesting to understand how this carries over to exceptional generalised geometry by developing a theory of generalised quotients.

Finally, returning to the AdS/CFT correspondence, one can consider  the structure of deformations. For example, in $\AdS5$ backgrounds, deforming the \HM{} structure while keeping $K$ fixed is equivalent to perturbing by chiral operators in the dual $\mathcal{N}=1$ SCFT. Requiring the new deformed structures to be integrable then restricts the form of the allowed deformations to marginal deformations. As we mentioned in the discussion of the moduli space of structures, one should take into account any extra isometries of the unperturbed background, as they define additional generalised Killing vectors. The quotient by these symmetries would then give the set of exactly marginal deformations in the SCFT. As we will show in a forthcoming paper~\cite{AGGPW15}, this gives the supergravity analogue of a well-known field theory result due to Green et al.~\cite{GKSTW10}.

\subsection*{Acknowledgements}

We would like to thank Charles Strickland-Constable and Mariana Gr\~{a}na for helpful discussions. AA is supported by an EPSRC PhD Studentship and COST Action MP1210. DW is supported by the STFC Consolidated Grant ST/L00044X/1, the EPSRC Programme Grant EP/K034456/1 ``New Geometric Structures from String Theory'' and the EPSRC Standard Grant EP/N007158/1 ``Geometry for String Model Building''. DW also thanks the Berkeley Center for Theoretical Physics at UC Berkeley for kind hospitality during the final stages of this work.  MP is partly supported by  ILP LABEX  (ANR-10-LABX-63) and the Idex SUPER (ANR-11-IDEX-0004-02).

\appendix

\section{Two results on normalisations and the supersymmetry conditions}
\label{app:lemma}

We first show that the $D=5$ normalisation condition $\kappa^2=c(K)$ is implied by the supersymmetry conditions for ESE spaces. Consider the set of generalised vectors of the form $V=fK$ where $f$ is an arbitrary function. Using the standard form of the generalised Lie derivative given in~\cite{CSW11}, we have 
\begin{equation}
   \Dorf_{fK} J_\alpha = f\Dorf_K J_\alpha 
      - \bigl[(\dd f \oadj K), J_\alpha\bigr] , 
\end{equation}
where $\oadj$ is the projection to the adjoint bundle $\oadj:E^*\otimes E\to\ad\tilde{F}$. Since $J_\alpha\cdot K=0$, we have $\tr \bigl((\dd f\oadj K)J_\alpha\bigr)=0$ and hence 
\begin{equation}
\begin{split}
   \epsilon_{\alpha\beta\gamma}\tr \bigl(J_\beta [\dd f\oadj K, J_\gamma]\bigr) 
      &= - \epsilon_{\alpha\beta\gamma}
            \tr \bigl((\dd f\oadj K) [J_\beta,J_\gamma]\bigr) \\
      &= - 2\kappa \tr \bigl((\dd f\oadj K)J_\alpha\bigr) \\
      &= 0 . 
\end{split}
\end{equation}
Thus 
\begin{equation}
   \mu_\alpha(fK) 
      = -\tfrac{1}{2} \epsilon_{\alpha\beta\gamma}
          \int_M f \tr (J_\beta \Dorf_K J_\gamma)
      = \lambda_\alpha \int_M f \kappa^2 ,
\end{equation}
where we have used the supersymmetry condition $\Dorf_K J_\alpha=\epsilon_{\alpha\beta\gamma}\lambda_\beta J_\gamma$. But we also have 
\begin{equation}
   \gamma(fK) = \int_M c(fK,K,K) = \int_M f c(K) . 
\end{equation}
Hence the moment map conditions~\eqref{eq:mm-gen} imply that 
\begin{equation}
   \int_M f \kappa^2 = \int_M f c(K) , \qquad \text{for all $f$}
\end{equation}
which implies the normalisation condition $\kappa^2=c(K)$. The analogous calculation in $D=4$ shows that the normalisation condition $\kappa^2=2\sqrt{q(K)}$ is similarly a consequence of the integrability conditions. 

Focussing again on $D=5$, for definiteness we set $\lambda_{1,2}=0$. We now show that for the action of $\GDiff_K$, that is those generalised diffeomorphisms that preserve $K$, the moment map conditions $\mu_+(V)=0$ are implied by the fixed-point conditions $\Dorf_K J_\alpha=\epsilon_{\alpha\beta\gamma}\lambda_\beta J_\gamma$, which read
\begin{equation}
\label{eq:DKJ-conds}
   \Dorf_K J_\pm = \pm \ii \lambda_3 J_\pm , \qquad
   \Dorf_K J_3 = 0 . 
\end{equation}
Acting on the first condition with $\Dorf_V$ we have 
\begin{equation}
   \ii \lambda_3 \Dorf_V J_+ 
      = \Dorf_V(\Dorf_K J_+) 
      = \Dorf_{\Dorf_VK} J_+ + \Dorf_K(\Dorf_V J_+)
      = \Dorf_K(\Dorf_V J_+) , 
\end{equation}
since  we have $\Dorf_VK=0$ for elements of the Lie algebra $\gdiff_K$. Substituting into the $\mu_+$ moment maps we have 
\begin{equation}
\begin{split}
   \mu_+(V) &\coloneqq - \ii \int_M \tr (J_3 \Dorf_V J_+) \\
      &= - \lambda_3^{-1} \int_M \tr (J_ 3 \Dorf_K\Dorf_V J_+) 
      = \lambda_3^{-1} \int_M \tr \bigl((\Dorf_KJ_ 3) (\Dorf_V J_+)\bigr)
      = 0 , 
\end{split}
\end{equation}
where we have used the second condition in~\eqref{eq:DKJ-conds}. 

\section{Flux quantisation, central charges and free energy}
\label{app:cc}

We briefly review the derivation of the central charge from~\cite{GGPSW09} and~\cite{GOV06}. The central charge $a$ is given in terms of the effective five-dimensional Newton's constant as~\cite{HS98}
\begin{equation}
   a = \frac{\pi}{8m^3 G_5} ,
\end{equation}
where $G_5$ in type IIB is given by 
\begin{equation}
\label{eq:G5-IIB}
   G^{-1}_{5,\text{IIB}} 
      = \frac{32\pi^2}{(2\pi\ell_s)^8g_s^2} \int_M \ee^{3\Delta'} \vol_5
      = \frac{32\pi^2}{(2\pi\ell_s)^8g_s^2} \int_M c(K) , 
\end{equation}
while for M-theory it is given by
\begin{equation}
\label{eq:G5-M}
   G^{-1}_{5,\text{M}} 
      = \frac{32\pi^2}{(2\pi\ell_{11})^9} \int_M \ee^{3\Delta} \vol_6
      = \frac{32\pi^2}{(2\pi\ell_{11})^9} \int_M c(K) . 
\end{equation}
The corresponding flux quantisation conditions are 
\begin{equation}
\begin{aligned}
   N &= \frac{1}{(2\pi\ell_s)^4g_s} \int_M \dd C \in \bbZ & && 
      & \text{type IIB} ,\\
   N_\Sigma &= \frac{1}{(2\pi\ell_{11})^3} \int_\Sigma \dd A \in \bbZ & && 
      & \text{M-theory} ,
\end{aligned}
\end{equation}
where $\Sigma$ is any four-cycle in $M$. From the five-dimensional part of Einstein's equations we note that $\dd C$ and $\dd A$ must both scale as the inverse AdS radius $m$. Defining the dimensionless volumes 
\begin{equation}
   V_5 = m^5 \int_M \ee^{3\Delta'} \vol_5 , \qquad
   V_6 = m^6 \int_M \ee^{3\Delta} \vol_6 , 
\end{equation}
we expect the scaling dependence 
\begin{equation}
   N \sim \frac{1}{m^4\ell_s^4g_s} V_5 , \qquad 
   N_\Sigma \sim \frac{1}{m^3\ell_{11}^3} V_6^{2/3} .
\end{equation}
More generally, as in~\cite{GGPSW09} and~\cite{GOV06}, one can solve explicitly for $\dd C$ and $\dd A$ in terms of the structure and find exact expressions for the flux quantisation. We also have
\begin{equation}
   a_{\text{IIB}} \sim \frac{1}{m^8\ell_s^8g_s^2} V_5 , \qquad
   a_{\text{IIB}} \sim \frac{1}{m^9\ell_{11}^9} V_6 . 
\end{equation}
Solving for $m$ then gives 
\begin{equation}
   a_{\text{IIB}} \sim \frac{N^2}{V_5} , \qquad 
   a_{\text{M}} \sim \frac{N_\Sigma^3}{V_6} ,
\end{equation}
and hence $a^{-1}$ scales as $\int_M c(K)$ in both cases. 

For M-theory $\AdS4$ backgrounds, we follow~\cite{GMPS14}. The free energy of the field theory is given by~\cite{EJM99}
\begin{equation}
   \mathcal{F} = \frac{\pi}{2m^2G_4} ,
\end{equation}
where the effective four-dimensional Newton's constant is
\begin{equation}
\label{eq:G4-M}
   G^{-1}_{4,\text{M}} 
      = \frac{32\pi^2}{(2\pi\ell_{11})^9} \int_M \ee^{2\Delta} \vol_7
      = \frac{32\pi^2}{(2\pi\ell_{11})^9} \int_M 2\sqrt{q(K)} . 
\end{equation}
The flux quantisation condition gives 
\begin{equation}
   N = \frac{1}{(2\pi\ell_{11})^6} \int_M \dd \tilde{A} \in \bbZ .
\end{equation}
Via the same scaling arguments as above, defining the dimensionless volume
\begin{equation}
   V_7 = m^7 \int_M \ee^{2\Delta} \vol_7 , 
\end{equation}
we find (the exact relations are given in~\cite{GMPS14})
\begin{equation}
   N \sim \frac{1}{m^6\ell_{11}^6} V_7 , \qquad 
   \mathcal{F} \sim \frac{1}{m^9\ell_{11}^9} V_7 ,
\end{equation}
so that solving for $m$ gives, as in~\cite{JKPS11},
\begin{equation}
   \mathcal{F} \sim \frac{N^{3/2}}{V_7^{1/2}} ,
\end{equation}
and hence $\mathcal{F}^{-2}$ scales as $\int_M \sqrt{q(K)}$.



\end{document}